\documentclass[12pt]{article}
\usepackage{titlesec}
\usepackage[margin=1.0in]{geometry}
\usepackage{enumitem}
\usepackage{comment}
\usepackage{authblk}
\usepackage[ruled,vlined]{algorithm2e}
\SetKwRepeat{Do}{do}{while}
\SetKwInput{KwInit}{Initialization}
\SetKwIF{If}{ElseIf}{Else}{if}{then}{else if}{else}{endif}
\usepackage{color}
\usepackage{setspace, float}
\usepackage{epsfig}
\usepackage{epstopdf}
\usepackage[section]{placeins}
\usepackage{graphicx}
\usepackage{subcaption}
\usepackage{amsmath}
\usepackage{amssymb}
\usepackage{bbm}
\usepackage{amsfonts}
\usepackage{mathrsfs}
\usepackage{multirow}
\usepackage{array}
\RequirePackage[numbers,sort&compress]{natbib}
\usepackage[
  colorlinks=true,
  linkcolor=blue,
  citecolor=blue,
  urlcolor=blue,
  backref=page
]{hyperref}
\usepackage{diagbox}
\usepackage{etoc, booktabs}
\AtBeginDocument{\etocdepthtag.toc{main}}

\newcommand{\pfend}{$\hfill\square$}

\newcommand{\indic}{\boldsymbol{\mathbbm{1}}}
\newcommand{\e}{\mathbf{e}}

\newcommand{\ZZ}{\mathbb{Z}}
\newcommand{\PP}{\mathbb{P}}
\newcommand{\EE}{\mathbb{E}}
\newcommand{\RR}{\mathbb{R}}

\newcommand{\CC}{\mathbb{C}}
\newcommand{\cum}{\mathrm{Cum}}

\newcommand{\nor}{\mathcal{N}}
\newcommand{\normal}[2]{\mathcal{N}\left(#1,#2\right)}

\renewcommand{\i}{\mathrm{i}}
\newcommand{\re}{\mathrm{Re}}
\newcommand{\im}{\mathrm{Im}}

\newcommand{\thres}{\Delta_f}

\newcommand{\kf}{\kappa_{f}}
\newcommand{\cmplt}{^\texttt{c}}

\newcommand{\kt}{\kappa_{\Theta}}

\def\bigO#1{\mathcal{O}\left(#1\right)} 
\def\bigOP#1{\mathcal{O}_\pr\left(#1\right)}

\newcommand{\converge}[1][]{\xrightarrow{#1}}

\newcommand{\dconverge}{\converge[d]}

\newcommand{\wmj}{\mathcal{W}_m(j)}

\newcommand{\thetahatdb}{\widetilde\Theta}

\newcommand{\tnf}{{\cal T}_n}

\newcommand{\vmnf}{{\cal V}_{m,n}}

\newcommand{\Bcal}{\mathcal{B}}

\newcommand{\M}{\mathcal{M}}

\newcommand{\lf}{{\mathcal{L}}}

\newcommand{\Fcal}{\mathcal{F}}

\newcommand{\thickhline}{
    \noalign {\ifnum 0=`}\fi \hrule height 1pt
    \futurelet \reserved@a \@xhline
}

\newcommand{\vertiii}[1]{{\left\vert\kern-0.25ex\left\vert\kern-0.25ex\left\vert #1 
    \right\vert\kern-0.25ex\right\vert\kern-0.25ex\right\vert}}

\newcommand{\maxnorm}[1]{{\left\vert\kern-0.25ex\left\vert #1 \right\vert\kern-0.25ex\right\vert}_{\max}}
\newcommand{\infnorm}[1]{{\left\vert\kern-0.25ex\left\vert #1 \right\vert\kern-0.25ex\right\vert}_{\infty}}

\def\bigOP#1{\mathcal{O}_{\PP}\left(#1\right)}

\DeclareMathOperator*{\argmin}{arg\,min}
\DeclareMathOperator*{\esup}{ess\,sup}
\DeclareMathOperator*{\cov}{Cov}
\DeclareMathOperator*{\var}{Var}
\DeclareMathOperator*{\pvar}{PVar}

\DeclareMathOperator*{\trace}{trace}

\newtheorem{theorem}{Theorem}[section]
\newtheorem{assumption}{Assumption}[section]
\newtheorem{lemma}{Lemma}[section]
\newtheorem{proposition}{Proposition}[section]

\newtheorem{remark}{Remark}[section]

\renewcommand*{\backref}[1]{}
\renewcommand*{\backrefalt}[4]{
  \ifcase #1 %
    (Not cited.)
  \or        %
    (Cited on page #2.)
  \else      %
    (Cited on pages #2.)
  \fi}
\providecommand{\keywords}[1]
{
{  \small	
  {\textit{Keywords---}} #1}
}
\begin{document}
\allowdisplaybreaks
\numberwithin{equation}{section}
\title{Inference for High-Dimensional\\ Sparse Spectral Precision Matrices}
\author[1]{Navonil Deb$^*$$^\dagger$}
\author[2]{Younghoon Kim$^*$}
\author[1]{Sumanta Basu}
\affil[1]{Department of Statistics and Data Science, Cornell University}
\affil[2]{Department of Mathematics, University of Alabama}
\date{}
\maketitle
\def\thefootnote{$*$}\footnotetext{Equal contribution}
\def\thefootnote{$\dagger$}\footnotetext{Corresponding author. Email: \href{mailto:nd329@cornell.edu}{\texttt{nd329@cornell.edu}}}
\begin{abstract}
Statistical inference for the spectral precision matrix at a given frequency allows us to assess frequency-specific conditional relationships among the components of a stationary multivariate time series.
Compared with classical analogues, inference in the spectral domain is more challenging due to the absence of closed-form asymptotic variance expressions for complex-valued estimators, the limited asymptotic theory in high-dimensional settings, and the presence of truncation and smoothing biases in finite samples.
We construct a debiased complex graphical lasso estimator at any fixed frequency by leveraging the full likelihood structure of neighboring discrete Fourier transforms. 
Using asymptotic distributions for bilinear forms of stationary multivariate time series, we establish the joint asymptotic normality of the real and imaginary parts of the debiased estimator. 
Our main theoretical contributions include deriving a closed-form asymptotic covariance matrix for the real and imaginary parts, establishing a central limit theorem for bilinear forms of the underlying time series, and controlling smoothing and truncation biases in covariance estimation to ensure valid inference.
Simulation studies demonstrate reliable coverage and improved statistical power relative to the benchmark, while maintaining false discovery rates near the nominal level. An application to real fMRI data further reveals distinct patterns of functional connectivity across selected frequencies.
\end{abstract}


\section{Introduction}\label{sec:introduction}

Spectral precision matrix (SPM) 
is central to the spectral domain analysis of multivariate stationary time series. Statistical inference of SPM is crucial for quantifying uncertainty in partial spectral coherence (PSC), the spectral domain analogue of partial correlation, which captures conditional linear association among the component processes, and also constructs graphical models of the time series across time lags. This problem is particularly important in applications including functional connectivity analysis in neuroimaging \citep{ombao2022spectral,wodeyar2022structural,baek2021detecting}, the recovery of financial networks \citep{barigozzi2017network,barigozzi2019nets}, and the analysis of environmental systems \citep{dahlhaus2000graphical}.

Formally, we consider a $p$-dimensional weakly stationary and centered real-valued time series $ X_t = \big(X_t^{(1)},\ldots,\allowbreak X_t^{(p)}\big)^\top$, $t\in \ZZ$, with autocovariance at lag $h \in \ZZ$ denoted by $\Gamma(h) := \cov(X_t, X_{t- h})= \EE\left[X_t X_{t-h}^\top\right]$. The spectral density matrix and spectral precision matrix (SPM) at $\omega \in [-\pi, \pi]$  are defined as 
\begin{equation*}
    f(\omega) := \frac{1}{2\pi}\sum_{h= -\infty}^\infty \Gamma(h) e^{-\i h \omega}, \quad \text{and} \quad
    \Theta(\omega) := f(\omega)^{-1}
\end{equation*}
respectively. Under stationarity, $\Gamma(-h)=\Gamma(h)^{\top}$, implying that $f(\omega)$ is Hermitian and thus is $\Theta(\omega)$ whenever $f(\omega)$ is invertible. Analogous to a precision matrix capturing the conditional dependencies in Gaussian graphical models (GGM), the SPM $(\Theta(\omega))_{a,b}$, $a,b=1,\ldots,p$, at the frequency $\omega$ captures the frequency-specific linear association between component processes $a$ and $b$ after their linear dependence on the remaining component processes has been removed.

A uniform penalty is typically used across the off-diagonal entries for regularized estimation of $\Theta(\omega)$ in high dimensions \citep{fiecas2019spectral,dallakyan2022time,baek2023local}, although the entries and their estimation variability can be substantially heterogeneous across the edges $(a,b)$ and frequency $\omega$ -- making the type-I error control of tests of nullity a difficult problem. This heterogeneity is relevant in resting-state fMRI data where functional connectivity strength varies across the regions and physiological activities \citep{ombao2022spectral,salvador2005undirected,chang2010time}. Characterizing the heterogeneity in entrywise variability is essential for adaptive estimation as the penalties should be calibrated to the estimation variability across the entries and the frequencies \citep{deb2024regularized, cai2011adaptive, cai2016estimating}. For a complex-valued SPM, this task requires deriving and estimating the $2\times 2$ joint asymptotic covariance matrix of the real and imaginary components of the estimator's every entry. Unlike in GGM with i.i.d. real samples \citep{ren2015asymptotic,jankova2015confidence,ning2017general,jankova2018inference}, inference of SPM is particularly difficult due to the complex-valued and entrywise heterogeneous nature of the estimator, along with the truncation and smoothing biases induced by finite-sample spectral estimation.

Parameters in the spectral domain are customarily estimated using the discrete Fourier transform (DFT) $d_k\in \{j-m,\ldots,j+m\}$ at frequencies near $\omega_j = 2\pi j/n$, formally defined as the inner product of the observed time series with the corresponding Fourier basis function (formal definition in \eqref{eq:dft}). Statistical inference of PSC is well established in finite-dimensional settings based on the inverse of the averaged smoothed periodogram $\sum_{k = j-m}^{j+m} d_k d_k^*/(2m+1)$ and multitaper spectral estimators \citep{priestley1988spectral,anderson2011statistical,brockwell1991time, fiecas2010functional,schneider2016p,tugnait2019edge}. However, these methods do not address the inference problem of SPM and PSC in high dimensions, including the corresponding asymptotic guarantees and derivation of the entrywise asymptotic covariance matrix. As a result, finite dimensional setings lack expressions for entrywise confidence intervals and the variance of the centered estimator. Recently, \citet{krampe2025frequency} developed high-dimensional inference procedures for PSC by exploiting its connection to the coefficients obtained from nodewise regressions of the DFT. In contrast, likelihood-based estimators such as the complex graphical Lasso \cite[CGLASSO;][]{deb2024regularized} exploit the joint structure of the DFT across all $p$ columns of the SPM. Empirically, such an estimator achieves more efficient model-selection performance than nodewise regression of the DFT that treats each SPM column as a separate regression coefficient and estimates it using a plug-in ratio of the SPM estimators \citep[Section 5.3]{deb2024regularized}.

In this work, we make three contributions to this literature. First, we propose an inference method of SPM by debiasing the complex graphical lasso (deCGLASSO). For a fixed frequency $\omega$, deCGLASSO builds upon the CGLASSO estimator of \citet{deb2024regularized}, which estimates $\Theta(\omega)$, and then estimates and corrects the regularization bias of CGLASSO using the graphical lasso debiasing technique for i.i.d. samples \citep{jankova2015confidence}. In Theorem \ref{thm:asymp_normal}, we derive the asymptotic covariance matrix of each entry of the deCGLASSO estimator and establish joint asymptotic normality of its real and imaginary parts. %
The asymptotic variance of the real part of deCGLASSO has the same functional form as the asymptotic variance of entries of the debiased graphical lasso estimator for real-valued i.i.d. samples \citep{jankova2015confidence}, up to a normalization constant. However, in the spectral domain, the variances of the real and imaginary parts, and their covariance, are not apparent from the asymptotic variance expression for i.i.d. real-valued samples and require a separate derivation.

Second, we extend the CLT for bilinear forms of time series \citep{wu2018asymptotic} to a joint setting (Lemma \ref{lem:joint_clt}), and use it for proving Theorem \ref{thm:asymp_normal} with controlling both the regularization bias and the spectral approximation bias. The proof of Theorem \ref{thm:asymp_normal} (Appendix~\ref{pf:asymp_normal}) requires a central limit theorem (CLT) on linear combinations of bilinear forms of $X_t$'s components, which is not directly applicable from the proof of the asymptotic normality of the debiased graphical lasso with i.i.d. real samples, that invokes the Martingale CLT \citep{jankova2015confidence}.
Next, we propose a plug-in estimator of the entrywise asymptotic covariance matrix that enables hypothesis tests and construction of confidence regions for the entries of the SPM. Consistent estimation of this covariance matrix is more challenging than the asymptotic variance in the debiased graphical lasso with i.i.d. real samples \citep{jankova2015confidence} because finite-sample truncation and spectral smoothing incur additional approximation bias. We correct the smoothing bias in our plug-in estimator and prove its consistency in Theorem \ref{thm:consistency_variance}. We additionally construct a heteroskedasticity autocorrelation consistent (HAC) estimator \citep{newey1987hypothesis} that directly plugs in the underlying covariance and fourth-order cumulant terms with their empirical counterparts, rather than reducing them to products of estimated second-order spectral quantities. 
Building upon the entrywise deCGLASSO estimates and these covariance matrix estimators, we establish a joint confidence ellipsoid for the real and imaginary parts of $\left(\Theta(\omega_j)\right)_{a,b}$ for $a\ne b$.

In addition, the deCGLASSO estimator can be further used to test the nullity of individual SPM entries and to perform multiple tests of nullity across entries \citep{liu2013gaussian}. The resulting confidence regions achieve near-nominal coverage in simulations as demonstrated in Section \ref{subsubsec:coverage}. The deCGLASSO estimator retains the joint likelihood of the neighboring DFT around $\omega_j$. Empirical results in Section~\ref{subsubsec:fdr_power} show competitive FDR control and improved power in calibrated fixed-frequency comparisons with the benchmark method of PSC based on nodewise regression of the DFT \citep{krampe2025frequency}, suggesting the efficiency gain we expected. In Section~\ref{subsec:data}, an analysis of resting-state fMRI data from the HCP Young Adult S1200 release shows that the estimated functional connectivity matrices exhibit distinctive patterns across frequencies, and deCGLASSO yields more discoveries than the PSC benchmark based on nodewise regression of the DFT \citep{krampe2025frequency}.

The rest of the paper is organized as follows. Section \ref{sec:method} introduces the preliminary concepts of spectral analysis, followed by regularized estimation of the spectral precision matrices and debiasing of the estimators. Section \ref{sec:theory} provides the theoretical guarantees of the proposed method and variance estimators for standardization of each entry. Section \ref{sec:numerical} illustrates the finite-sample performance using both synthetic and resting-state fMRI data. Section \ref{sec:discussion} contains the discussion of future directions. The details of the proofs and additional simulation results are provided in Appendix \ref{pf:reg_bias}-\ref{app:additional_sim}. The codes used for this paper are available in our R package \texttt{cxreg} \citep{kim2026cxreg}.

\paragraph{Notation}  For $z\in \mathbb{C}$, $\re(z)$ and $\im(z)$ denote the real and imaginary parts of $z$, respectively. $z^{*}$ and $\overline{z}$ both denote its complex conjugate, and $|z|=\sqrt{\re(z)^2+\im(z)^2}$. For $p\in\mathbb{N}$, $\mathbf{e}_a$ denotes the $p$-dimensional basis vector with 1 at $a$th entry and 0 otherwise. 
For a complex vector $\mathbf{z}=(z_1.\ldots,z_p)\in\mathbb{C}^p$, $\re(\mathbf{z})$, $\im(\mathbf{z})$, $\overline{\mathbf{z}}$, and $\mathbf{z}^*$ denote their real parts, imaginary parts, conjugate, and conjugate transpose, respectively. For $q>0$, $\|\mathbf{z}\|_q:=(\sum_{a}|z_a|^q)^{1/q}$. 
For matrix $A\in\mathbb{C}^{p_1 \times p_2}$, $p_1,p_2\in\mathbb{N}$, $A^{\top}, \overline{A}, A^*$ denote its transpose, conjugate and conjugate transpose respectively. $A_{a,\cdot}$, $A_{\cdot,b}$, and $(A)_{a,b}$ represent its $a$th row, $b$th column, and an entry at $a$th row and $b$th column respectively. If $A$ is a square matrix, $\mathrm{trace}(A)$ and $\mathrm{det}(A)$ denote its trace and determinant, respectively. $A^-$ denotes the matrix obtained from setting the diagonals of $A$ to zero. 
$\mathcal{H}_{++}^{p}:=\{A\in\mathbb{C}^{p\times p}:A=A^{*},A >0\}$. We also denote $|||A|||_{\infty}:=\max_{a}\sum_{b}|(A)_{a,b}|$, $||| A |||_1:=\max_{b}\sum_{a}|(A)_{a,b}|$, $\|A\|:=\max_k\textrm{eig}_k(A^{*}A)$, $\|A\|_{F}:=\sqrt{\textrm{trace}(A^{*}A)}$, and $\|A\|_{\infty}=\max_{a,b}|(A)_{a,b}|$. I
For matrices $A,B\in\mathbb{C}^{p_1 \times p_2}$, $A\otimes B$ denotes the Kronecker product of $A$ and $B$. 
For sequences $(x_n)_{n\in\mathbb{N}}$ and $(y_n)_{n\in\mathbb{N}}$, we denote $x_n \succeq y_n$ if there is a universal constant $c > 0$, independent of the dimension and model parameters, such that $x_n \geq c y_n$ for all $n\in\mathbb{N}$. $x_n \preceq y_n$ and $x_n = \mathcal{O}(y_n)$ are defined analogously. $x_n = o(y_n)$ denotes $x_n/y_n \to 0$ as $n \to \infty$. $x_n \asymp y_n$ denotes that both $x_n \succeq y_n$ and $x_n \preceq y_n$ hold. 
For r.v.’s $X_n$ and sequence $a_n$, $X_n=\bigOP{a_n}$ means $X_n/a_n$ is bounded in probability, i.e. for every $\delta > 0$ there exists $c_\delta$ and $n_\delta$ such that $\PP\left( \frac{|X_n|}{a_n} > c_\delta \right) < \delta$ for all $n > n_\delta$. $X_n = o_\PP(Y_n)$ means $X_n/Y_n \xrightarrow{\PP} 0$. For complex valued random variables $Z_1, Z_2$, we denote the cumulant as $\cum(Z_1, Z_2) = \EE[(Z_1 - \EE[Z_1]) (Z_2 - \EE[Z_2])]$, the covariance as $\cov(Z_1, Z_2) = \EE[(Z_1 - \EE[Z_1]) \overline{(Z_2 - \EE[Z_2])}]$, $\var(Z_1) = \EE[|Z_1 - \EE[Z_1]|^2]$ and the pseudovariance as $\pvar(Z_1) = \EE[(Z_1 - \EE[Z_1])^2]$.

\section{Method}\label{sec:method}

We start by providing the background on spectral domain objects used in the paper, followed by sparse estimation of the SPM to obtain the CGLASSO estimator and the deCGLASSO estimator for inference.

Consider the grid of Fourier frequencies $\omega_j = 2\pi j/n$, where $n$ is the number of samples, $j \in F_n := \left\{ -\left\lfloor(n-1)/2\right\rfloor, \ldots, \left\lfloor n/2 \right\rfloor \right\}$ for a numerical purpose. Given $n$ consecutive observations $X_1,\ldots,X_n$, the discrete Fourier transform (DFT) at frequency $\omega_j$ is defined as 
\begin{equation}\label{eq:dft}
    d_j:= d(\omega_j) = \frac{1}{\sqrt{2\pi n}} \sum_{t=1}^n X_t e^{-\i t \omega_j}.
\end{equation}
We denote $B_n := \{-\lfloor(n-1)/2 \rfloor, 0, \lfloor n/2 \rfloor \}$ as the set of \textit{boundary frequencies}. For $j\in F_n$, we denote the set of its $2m+1$ nearby frequencies by $\wmj := \{k \in F_n: |k - j| \le m\}$, where the frequency indices are evaluated modulo $n$, since $k-j$ may fall outside $F_n$. Following \citet[Theorem~11.7.1]{brockwell1991time}, \citet[Theorem~4.4.1]{brillinger2001time}, and under regularity conditions on the time series $X_t$, the DFT $d_j$ for $j\in F_n \setminus B_n$ converges weakly to a mean-zero complex Gaussian random vector with covariance matrix $f(\omega_j)$ as $n \to \infty$. Moreover, under regularity conditions (Assumption \ref{asn:summable}), the spectral density is Lipschitz smooth and hence $f(\omega_k) \approx f(\omega_j)$ when $k$ is close to $j$. Furthermore, for stationary $\{X_t\}_{t\in[n]}$, the coordinates of $d_k$ and $d_j$ are asymptotically uncorrelated (see Lemma \ref{lem:dft_cumulant} for details). Therefore, the collection $\{d_k\}_{k\in \wmj}$ can be viewed as approximately and asymptotically independent complex Gaussian random vectors with complex-valued covariance matrix $f(\omega_j)$. This approximation enables the construction of estimators for $f(\omega_j)$ and $\Theta(\omega_j)$ for $k\in F_n$ by treating $d_k$, $k \in \wmj$ as the spectral domain data, using methods analogous to those developed for i.i.d. real-valued data, such as averaged smoothed periodogram, kernel-based estimators and other spectral domain estimators derived from them \citep{brockwell1991time, brillinger2001time, priestley1988spectral, hannan2009multiple}.

The averaged periodogram estimator at $\omega_j$ is denoted by
$$ \hat f(\omega_j) \equiv \hat f_j = \frac{1}{2m+1}\sum_{k \in \wmj} d_k d_k^*. $$
For simplicity, we use a rectangular kernel for averaging the periodograms. Other kernels can also be used \citep{wu2018asymptotic, basu2023graphical}. In the classical asymptotic regime, $m = \bigO{n^{0.8}}$ is a standard choice of the bandwidth for obtaining the optimal mean squared error \citep{delgado1996optimal, nordman2007empirical}. In contrast, taking $m = \bigO{\sqrt{n}}$ \citep{qu2011test, bohm2009shrinkage} in high-dimensional regimes is known in the literature to ensure that $\hat f_j$ is consistent for $f_j$. We adopt the latter choice of bandwidth for our empirical analysis, and also an adaptive choice based on generalized cross-validation \citep{ombao2001simple} in Section \ref{sec:numerical}.

\subsection{deCGLASSO estimator}\label{sec:debiased_cglasso}

As proposed in \citet{deb2024regularized}, the complex graphical lasso (CGLASSO) estimator at Fourier frequency $\omega_j$, denoted by $\hat\Theta_j \equiv \hat\Theta(\omega_j)$ is the minimizer of the optimization problem
\begin{equation}\label{eq:cglasso}
    \min_{\Theta \in \mathcal{H}^p_{++}} \left\{L(\Theta) + \lambda P(\Theta)\right\} = \min_{\Theta \in \mathcal{H}^p_{++}} \left\{-\log \det \Theta + \trace(\hat f_j\Theta) + \lambda \|\Theta^-\|_{1}\right\},
\end{equation}
where $\lambda > 0$ is a regularization parameter. We use the shorthands $L(\Theta) := -\log \det\Theta + \trace(\hat f_j \Theta)$ and $P(\Theta) :=\|\Theta^-\|_{1} = \sum_{a\ne b}|\Theta_{a,b}|$. The Karush–Kuhn–Tucker (KKT) condition \citep{boyd2004convex} corresponding to \eqref{eq:cglasso} is 
\begin{equation}\label{eqn:kkt}
    \nabla L(\hat\Theta_j) + \lambda \\\eta(\hat\Theta_j) = -\hat\Theta_j^{-1} + \hat f_j + \lambda \hat \Psi = 0,
\end{equation}
where $\eta(\cdot)$ is the sub-gradient of $P(\cdot)$, and $\hat\Psi = (\hat\psi_{a,b})_{a,b=1}^p$ is denoted by 
\begin{equation*}
    \hat\psi_{a,b} 
    = \begin{cases}
    ~\quad 0 & \text{if } a = b,\\
    \quad \dfrac{(\hat\Theta_j)_{a,b}}{|(\hat\Theta_j)_{a,b}|} & \text{if } a \neq b\ \text{and } (\hat\Theta_j)_{a,b} \neq 0,\\
    \in \{z\in \CC: |z|\leq 1\} & \text{if } a\neq b\ \text{and } (\hat\Theta_j)_{a,b} = 0.
\end{cases}
\end{equation*}

The debiased complex graphical lasso (deCGLASSO) estimator is a frequency-domain generalization of the debiased GLASSO estimator for i.i.d. real-valued samples \citep{jankova2015confidence,jankova2018inference}. In the low-dimensional setting and under no regularization, i.e., $\lambda = 0$, $\hat \Theta_j$ is the inverse of $\hat f_j$. However, in high dimensions, $\lambda\hat\Psi$ in the KKT condition \eqref{eqn:kkt} induces a \emph{regularization bias}. The debiased estimator, denoted by $\thetahatdb_j$, is thus the solution satisfying $\nabla L(\thetahatdb_j)\approx0$. Therefore the first-order Taylor's expansion of $\nabla L(\thetahatdb_j)$ around $\hat\Theta_j$ satisfies
\begin{displaymath}
    \nabla L(\hat\Theta_j) + \nabla^2 L(\hat\Theta_j)(\thetahatdb_j-\hat\Theta_j) 
    = \nabla L(\hat\Theta_j) + \lambda \eta(\hat\Theta_j) = 0.
\end{displaymath}
The deCGLASSO estimator $\thetahatdb_j$ is thus obtained as
\begin{equation}\label{eqn:debiased_cglasso}
    \thetahatdb_j = \hat\Theta_j + \lambda \hat\Theta_j \hat\Psi \hat\Theta_j = 2\hat\Theta_j - \hat\Theta_j \hat f_j\hat\Theta_j.
\end{equation}
We note that pre- and post-multiplying $\hat\Theta_j$ to \eqref{eqn:kkt} yields $\hat{\Theta}_j - \hat{\Theta}_j \hat f_j \hat{\Theta}_j = \lambda \hat{\Theta}_j \hat{\Psi}\hat{\Theta}_j$ that can be regarded as an additional term for bias correction. In the derivation of the asymptotic distribution of the $\thetahatdb_j$ (Theorem \ref{thm:asymp_normal}), we will show that the bias correction can be controlled in appropriate regimes. Specifically, we let $W_j := \hat f_j - f_j$. By rearranging the terms in \eqref{eqn:debiased_cglasso} and using \eqref{eqn:kkt} again, we have
\begin{equation}\label{eq:bias_variance_decomposition}
    \thetahatdb_j - \Theta_j = \hat\Theta_j + \lambda \hat\Theta_j \hat\Psi\hat\Theta_j  - \Theta_j = -\Theta_j W_j \Theta_j + R_{n,p},
\end{equation}
where 
\begin{equation}\label{eq:remainder}
    R_{n,p} := -(\hat\Theta_j - \Theta_j) W_j \Theta_j - (\hat\Theta_j \hat f_j - I_p)(\hat\Theta_j - \Theta_j).
\end{equation}
$R_{n,p}$ is the \textit{regularization bias}. In Proposition \ref{prop:reg_bias}, we show along \citet[Lemma 1]{jankova2015confidence} that $R_{n,p}$ is asymptotically small, and the leading term $-\Theta_j W_j \Theta_j$ determines the entry-wise asymptotical distribution of $\thetahatdb_j - \Theta_j$.

In Theorem \ref{thm:asymp_normal}, we show that $T_j^{(a,b)} := \sqrt{2m+1}(\tilde{\Theta} - \Theta)$ is asymptotically distributed as the random variable $\zeta_1 + \i \zeta_2$, where $(\zeta_1, \zeta_2)^\top$ is a bivariate Gaussian random variable with $\var(\zeta_1) = \left[(\Theta_j)_{a,a} (\Theta_j)_{b,b} + \re((\Theta_j)_{a,b}^2)\right]/2, \var(\zeta_2) = \left[(\Theta_j)_{a,a} (\Theta_j)_{b,b} - \re((\Theta_j)_{a,b}^2)\right]/2$ and $\cov(\eta_1, \eta_2) = \im((\Theta_j)_{a,b}^2)/2$. In Section \ref{subsec:asymptotic_variance_covariance}, we develop consistent estimators for these population parameters, and use them to construct confidence regions and tests for nullity of the SPM entries.

\section{Theoretical results}
\label{sec:theory}

In this section, we derive the asymptotic normality along with the asymptotic covariance of the real and imaginary parts of the deCGLASSO estimator, and provide consistent estimators of the asymptotic variance, followed by entry-wise confidence intervals. First, we state two assumptions required for controlling the estimated bias of the CGLASSO estimator and proving its asymptotic normality.

\begin{assumption}[First order summability]\label{asn:summable}
    $\{X_t\}_{t\in [n]}$ is a stationary, Gaussian and centered time series with ACF $\Gamma(\cdot)$ satisfying
    \begin{equation}\label{eq:summable}
        \sum_{h=0}^\infty h \|\Gamma(h)\| < \infty.
    \end{equation}
\end{assumption}

Assumption \ref{asn:summable} implies that the spectral density is finite at every frequency. We denote the following two measures of stability
$$ \vertiii{f} = \esup_{\omega \in [-\pi, \pi]} \|f(\omega)\|, \quad \vertiii{\Theta} = \esup_{\omega \in [-\pi, \pi]} \|\Theta(\omega)\|. $$
that appear in the theoretical analysis of \citep{deb2024regularized}. Assumption \ref{asn:summable} guarantees that $f$ is Lipschitz continuous with Lipschitz constant $\lf/(4\pi)$, where $\lf := 4 \sum_{h =1}^\infty h \|\Gamma(h)\|$. In addition, Assumption \ref{asn:summable} guarantees the finiteness of the \textit{the measure of stability} $\M := \max\{\vertiii{f}, \vertiii{\Theta}\}$.

Next, we define some model parameters that appear in our theoretical results.

\paragraph*{Approximation bias} Define a model dependent parameter
\begin{equation}\label{eq:tnf}
     \tnf := \frac{1}{2\pi} \sum_{|h|>n} \| \Gamma(h) \|,
\end{equation}
that captures the strength of the temporal and contemporaneous memory at the tail of the ACFs \citep{sun2018large}. Denote
\begin{equation}\label{eq:approximation_bias}
    \vmnf := \tnf + \frac{m}{n}\lf
\end{equation}
as the \textit{approximation bias} that appears in the error bounds of $\hat f_j$ and $R_{n,p}$.

\paragraph*{Sparsity parameters} The edge set of $\Theta_j$ is denoted by $E := E(\Theta_j) = \{(a,b)\in [p]^2: a\ne b, (\Theta_j)_{a,b} \neq 0\}$ (excluding the diagonals, i.e., self-loop edges), and the augmented edge set by $S:= S(\Theta_j) = E(\Theta_j)\cup \{(1, 1),\cdots, (p,p)\}$. The two sparsity parameters required for our analysis are the \textit{number of edges} $s := |E|$ and the \textit{maximum degree} $d := \max_{a \in [p]}\left|\{b \in [p]:\ (\Theta_j)_{a,b} \neq 0\}\right|$.

\paragraph*{Model parameters} Consider the function $g(\Theta) = \log \det \Theta,\ \Theta \succ0$. Following \citet{boyd2004convex}, the Hessian of $g$ is given by 
\begin{equation*}
    \Upsilon := \nabla^2 g(\Theta_j) = (\Theta_j)^{-1}\otimes (\Theta_j)^{-1} = f \otimes f \in \CC^{p^2\times p^2},
\end{equation*}
and is indexed with the edges such that for $e_1, e_2 \in [p]^2$, the entries $(\Upsilon)_{e_1, e_2}$ are of the form $\frac{\partial^2 g}{\partial \Theta_{e_1}\partial \Theta_{e_2}}(\Theta)$ evaluated at $\Theta = \Theta_j$. Similarly, $\Upsilon_{S_1,S_2}$ are defined for the $|S_1|\times|S_2|$ matrix with rows and columns of $\Upsilon$. In addition, define  $\kf := \vertiii{f}_\infty$ and $\kt := \vertiii{(\Upsilon)_{S,S}}_\infty$.

We now state an assumption to be used in our non-asymptotic analysis.

\begin{assumption}[Incoherence] \label{assumption:incoherence}
There exists $\gamma \in (0,1]$ such that 
$$\max_{e\in S\cmplt} \|\Upsilon_{e,S} \left(\Upsilon_{S,S} \right)^{-1} \|_1 \leq 1-\gamma.$$
\end{assumption}

Assumption \ref{assumption:incoherence} is standard in the literature of the GLASSO estimator, ensuring that the non-edge indices of the Hessian have limited influence on the edge indices \citep{zhao2006model, ravikumar2011high}. We additionally denote 
$C_\gamma := 1+ \frac{8}{\gamma}$, required in the results.

\subsection{Bounds on regularization bias}
\label{subsec:regularization bias}

We show that the regularization bias of the deCGLASSO estimator is bounded in an appropriate regime. For $A > 0$, we denote,
\begin{equation}\label{eq:threshold}
    \thres := \M\sqrt{\frac{A\log p}{m}} + \vmnf,
\end{equation}
that appears in the error bound of $R_{n,p}$. The following proposition states that $R_{n,p}$ is bounded with high probability in the appropriate regime.

\begin{proposition}[Control on the regularization bias]\label{prop:reg_bias}
    Let $\{X_t\}_{t\in[n]}$ be a stationary time series satisfying Assumptions \ref{asn:summable} and \ref{assumption:incoherence}. For $\kf, \kt \ge 1$, assume that $n$ satisfies $n \succsim \lf \M^3 \log p$ and $\tnf \le 1/(4\M)$. For any $A>0$, $\thres$ in \eqref{eq:threshold} satisfying $d \thres \leq [6 \kt^2 \kf^3 C_\gamma]^{-1}$, with the choice of the regularization parameter $\lambda = \frac{8}{\gamma}\thres$ and the bandwidth such that $m \succsim \M^2 \log p$ and $ m \le n/(4\M \lf)$, there exist constants $c, c'>0$ such that $R_{n,p}$ in \eqref{eq:remainder} satisfies 
    \begin{displaymath}
        \|R_{n,p}\|_\infty =  \bigO{\frac{\M\kt^2 d\thres ^2}{\gamma^2}}, 
    \end{displaymath}
    with probability greater than $1 - c \exp(-(c'A - 2)\log p)$.
\end{proposition}

Proposition \ref{prop:reg_bias} establishes a non-asymptotic error bound on the regularization bias of deCGLASSO. If $\M$ and $\kt$ are asymptotically bounded above and $\gamma$ is asymptotically bounded below, then the sparsity assumption becomes $d\thres^2 = o(1)$, which shows a similar result in \citet[Lemma 1]{jankova2015confidence}. A sufficient condition is $d \sqrt{\log p/m} = o(1)$ and $d m/n = o(1)$. In particular, if the bandwidth satisfies $m \asymp n^\xi$ for some $\xi \in (0,1)$ and $n \succsim (\M\lf)^{-(1-\xi)}$, then 
\begin{displaymath}
d = o \left(\min\left\{ \frac{n^{\xi/2}}{\sqrt{\log p}}, n^{1-\xi} \right\}\right)
\end{displaymath}
ensures that $\|R_{n,p}\|_\infty = \bigOP{d \log p/m}$, which is the same as the upper bound for the bias of the debiased GLASSO estimator with $m$ i.i.d. real-valued settings \citep{jankova2015confidence}. $R_{n,p}$ vanishes with high probability for $d\log p/m \to 0$.

\subsection{Asymptotic distribution}\label{subsec:asymptotic_distribution}

The next result is the building block for our proposed inference framework. We fix the coordinate $(a,b) \in [p]^2$, and derive the asymptotic distribution of $(\thetahatdb_j)_{a,b}$. We denote $T_j^{(a,b)} := \sqrt{2m+1}\left((\thetahatdb_j)_{a,b} - (\Theta_j)_{a,b}\right)$.

\begin{theorem}[Asymptotic normality]\label{thm:asymp_normal}
    Let $\{X_t\}_{t\in[n]}$ be a stationary time series satisfying Assumption \ref{asn:summable} and \ref{assumption:incoherence}. Assume that the conditions of Proposition \ref{prop:reg_bias} are satisfied. Furthermore, let $ \vmnf = o\left(\frac{1}{\sqrt{m}\M^2}\right)$ and $d\thres ^2 = o\left(\frac{\gamma^2}{\M \kt^2}\right)$. Then
    \begin{displaymath}
        T_j^{(a,b)} = \zeta_1 + \i \zeta_2 + o_\PP(1)
    \end{displaymath}
    where $(\zeta_1, \zeta_2)^\top$ is a bivariate Gaussian random variable with $\EE[\zeta_1]=\EE[\zeta_2] = 0$ and
    \begin{align}
        \EE[\zeta_1^2] = & \frac{1}{2}\left[(\Theta_j)_{a,a} (\Theta_j)_{b,b} + \re((\Theta_j)_{a,b}^2)\right], \label{eq:var_real} \\
        \EE[\zeta_2^2] = & \frac{1}{2}\left[(\Theta_j)_{a,a} (\Theta_j)_{b,b} - \re((\Theta_j)_{a,b}^2)\right], \label{eq:var_imaginary} \\
        \EE[\zeta_1 \zeta_2] = & \frac{1}{2} \im((\Theta_j)_{a,b}^2). \label{eq:var_cov} 
    \end{align}
\end{theorem}

The proof is based on a central limit theorem of the general bilinear form of a stationary $\{X_t\}_{t\in[n]}$ (Lemma \ref{lem:cmplx_clt}). Note that it is different from the statement in i.i.d. real-valued samples in \citet{jankova2015confidence} since the DFTs are only \emph{asymptotically independent} even in fixed-dimension cases. Therefore, we will use the results of \citet{liu2010asymptotics,wu2018asymptotic} for entry-wise asymptotic normality of the averaged periodogram estimator, combining with the fact that the deCGLASSO estimator is a linear combination of the entries of $W_j$.

The condition $ \vmnf = o(1/(\sqrt{m}\M^2))$ ensures that the approximation bias remains sufficiently small and does not dominate the stochastic term $\sqrt{\log p/m}$, which is satisfied under the condition $n \succeq \lf \M^2 m^{3/2}$. The condition $d\thres^2 = o(\gamma^2/(\M\kt^2))$ also implies that $R_{n,p}$ arising from regularization bias vanishes, by the result of Proposition \ref{prop:reg_bias}. If $\gamma, \M, \kt$ are asymptotically bounded, this condition is equivalent to $d\log p/m = o(1)$ and $dm^2/n^2 = o(1)$.

The covariance expressions in \eqref{eq:var_real}--\eqref{eq:var_cov} generalize the real-valued i.i.d. setting, in which the variance formula for entries of the debiased GLASSO estimator \citep{jankova2015confidence} has the same functional form as the variance of the real part in \eqref{eq:var_real}, up to a multiplicative normalization constant. A similar structure also appears in the asymptotic variance of lag-window estimators of $f_j$ \citep{wu2018asymptotic}. The asymptotic normality in Theorem~\ref{thm:asymp_normal} can be interpreted by noting that the DFTs $\{d_k : k \in \wmj\}$ behave asymptotically as independent, zero-mean complex Gaussian vectors with covariance matrix $f_j$. As shown in the Appendix equation \eqref{eq:decomp}, the regularization bias-free term $\Theta_j W_j \Theta_j$ in \eqref{eq:bias_variance_decomposition} is asymptotically equivalent to an average of outer products of the scaled vectors $\Theta_j d_k$, which are asymptotically independent and follow a zero-mean complex Gaussian distribution with covariance matrix $\Theta_j$. In the real-valued i.i.d. setting for $Z\sim\normal{0}{\Omega}$, $\operatorname{Var}(Z_aZ_b)=\Omega_{a,a}\Omega_{b,b}+\Omega_{a,b}^2$ by the Gaussian fourth-moment identity \citep{isserlis1918formula} -- yielding the entrywise asymptotic variance for the debiased graphical lasso estimator \citep{jankova2015confidence}.

\subsection{Asymptotic variance and covariance estimators}\label{subsec:asymptotic_variance_covariance}

Since the variances of the real and imaginary parts of $\thetahatdb_j$ in Theorem \ref{thm:asymp_normal} and their covariance are unknown, we estimate these quantities so that we leverage them for constructing entry-wise confidence intervals and performing hypothesis tests for the entries of the SPM.

Unlike i.i.d. real-valued sample cases \citep{jankova2015confidence, jankova2018inference}, however, $\Theta_j d_k$ admits heterogeneous covariance matrices across for varying $k$. Even though this issue of heterogeneity is asymptotically marginal for proving Theorem \ref{thm:asymp_normal}, the truncation bias remains for the scaled DFTs. To circumvent this truncation bias, we construct estimators for the asymptotic variance and covariance of the real and imaginary parts of $\thetahatdb$ by directly calculating the variance and pseudovariance of the averaging terms $\Theta_j d_k d_k^* \Theta_j$ in $T_j^{(a,b)}$.

We now introduce two sets of consistent estimators for the entry-wise variance and covariance of the real and imaginary parts of $\thetahatdb_j$. First, we propose a plug-in estimator tailored to the complex-valued setting \citep{dwivedi2011test}, and refer to the resulting estimator as the plug-in estimator. Second, we present a heteroskedasticity and autocorrelation consistent (HAC) variance estimator \citep{newey1987hypothesis} that exploits the cumulant-based representation of the variance and pseudovariance of $\Theta_j^* W \Theta_j^*$.

\paragraph*{Plug-in estimator} 

We denote the entry-wise asymptotic variance and the pseudovariance of each term in the difference as 
\begin{equation}\label{eq:var_pseudovar_scaled_diff_pop}
    (\sigma_j^2)^{(a,b)} := \var \left((\Theta_j)_{\cdot,a}^* W (\Theta_j)_{\cdot,b} \right), \ 
    (\delta_j^2)^{(a,b)} := \pvar \left((\Theta_j)_{\cdot}^* W (\Theta_j)_{\cdot,b} \right).
\end{equation}
Denote $\hat\Phi_{j,k}^{(a,b)} := (\hat\Theta_j)_{\cdot,a}^* \hat f_k (\hat\Theta_j)_{\cdot,b}$ and $\hat\Xi_{j,k}^{(a,b)} := (\hat\Theta_j)_{\cdot,a}^\top \hat f_k (\hat\Theta_j)_{\cdot,b}$ for $j \in F_n$ and $k \in \wmj$. The estimators of the variance and pseudovariance in \eqref{eq:var_pseudovar_scaled_diff_pop} are respectively
\begin{align*}
    (\hat\sigma_j^2)^{(a,b)} 
    = \frac{1}{(2m+1)^2} \left( \sum_{k\in \wmj} \hat \Phi_{j,k}^{(a,a)} \hat \Phi_{j,k}^{(b,b)} + \sum_{k_1, k_2 \in \wmj} \hat\Xi_{j,k_1}^{(a,b)} \overline{\hat\Xi_{j,k_1}^{(a,b)}} \indic\left\{k_1 + k_2 \in n \ZZ\right\} \right), \\
    (\hat\delta_j^2)^{(a,b)} 
    = \frac{1}{(2m+1)^2} \left( \sum_{k\in \wmj} \left(\hat \Phi_{j,k_1}^{(a,b)}\right)^2 + \sum_{k_1, k_2 \in \wmj} \overline{\hat \Xi_{j,k_1}^{(a,a)}} \hat \Xi_{j,k_1}^{(b,b)} \indic\left\{k_1 + k_2 \in n \ZZ\right\} \right). 
\end{align*}
The plug-in estimator of the $2\times 2$-dimensional joint covariance matrix of the real and imaginary parts of $\thetahatdb$ is denoted by $\hat\Sigma_j^{(a,b)}$, where
\begin{align}
    (\hat\Sigma_j^{(a,b)})_{1,1} = & \frac{1}{2}\left[(\hat \sigma_j^2)^{(a,b)} + \re\left((\hat \delta_j^2)^{(a,b)}\right)\right], \label{eq:var_re_plugin} \\
    (\hat\Sigma_j^{(a,b)})_{2,2} = & \frac{1}{2}\left[(\hat \sigma_j^2)^{(a,b)} - \re\left((\hat \delta_j^2)^{(a,b)}\right)\right], \label{eq:var_im_plugin} \\
    (\hat\Sigma_j^{(a,b)})_{1,2} = & \frac{1}{2} \im\left((\hat \delta_j^2)^{(a,b)}\right). \label{eq:cov_re_im_plugin} 
\end{align}
The plug-in estimator is directly motivated by the population-level variance and pseudovariance expressions, and thus preserves a clear connection to the underlying theoretical quantities. However, this approach implicitly relies on accurate estimation of $\hat f_k$ across frequencies. It may be sensitive to estimation errors. In addition, the plug-in estimator involves approximating higher-order dependence through products of second-order quantities, which can introduce additional variability. Despite these limitations, the plug-in estimator serves as a natural benchmark due to its simplicity.

\paragraph{HAC-type estimator}
Denote $\hat Y_{j,k}^{(a)} := (\hat\Theta_j)_{\cdot,a}^{*}d_k$ and $\hat y_{j,k}^{(a,b)} := \hat Y_{j,k}^{(a)} \overline{\hat Y_{j, k}^{(b)}} - (\hat\Theta_{j})_{\cdot,a}^*\hat f_{k}(\hat\Theta_{j})_{\cdot,b}$, and $W_{k_1, k_2} := \indic(k_1 = k_2) + \indic(k_1 + k_2 \in n\ZZ)$. Another set of estimators for the variance and pseudovariance are respectively
\begin{align}
    (\hat \sigma_j^2)^{(a,b); \mathrm{HAC}}
    := \frac{1}{(2m+1)^2} \sum_{k_1, k_2 \in \wmj}^m W_{k_1, k_2} \hat y_{j,k_1}^{(a,b)} \overline{\hat y_{j,k_2}^{(a,b)}} , \label{eq:variance_HAC} \\
    (\hat \delta_j^2)^{(a,b); \mathrm{HAC}}
    := \frac{1}{(2m+1)^2} \sum_{k_1, k_2 \in \wmj}^m W_{k_1, k_2} \hat y_{j,k_1}^{(a,b)} \hat y_{j,k_2}^{(a,b)}, \label{eq:pseudovariance_HAC}
\end{align}
respectively. An HAC-type alternative estimator of the real-imaginary covariance matrix $(\hat\Sigma_j)^{(a,b); \mathrm{HAC}}$ can be defined analogously to \eqref{eq:var_re_plugin}--\eqref{eq:cov_re_im_plugin}. As discussed in Appendix \ref{sec:consistency_hac}, this estimator is constructed by replacing the fourth-order cumulants with their empirical counterparts. A key advantage of the HAC estimator is its ability to account for dependence across frequencies by appropriately weighting pairs of DFTs. In \eqref{eq:variance_HAC} and \eqref{eq:pseudovariance_HAC}, the weights $W_{k_1,k_2}$ can be generalized to kernel-based weights, yielding more flexible, data-driven estimators of the variance and pseudovariance, analogous to time-domain methods for handling heteroscedasticity \citep{andrews1991heteroskedasticity}. This generalization can provide increased robustness to mild non-stationarity and the resulting 
cross-frequency dependence.

Next we make a sufficient assumption for consistency of the HAC estimator.

\begin{assumption}[Finite 8\textsuperscript{th} moment]\label{asn:8th_moment}
    For $3 \le r \le 8$,
    $$\M_4 := \max_{a_1,\ldots, a_r\in [p]}\sum_{h_1, \ldots, h_{r-1} \in \ZZ} |h_\ell| \left|\cum\left(X_0^{(a_1)}, X_{h_1}^{(a_2)}, \ldots, X_{h_{r-1}}^{(a_r)}\right)\right| = \bigO{1}.$$ 
\end{assumption}

This assumption is similar to \citet[Equation 4.3.10]{brillinger2001time}.

\begin{theorem}[Consistency of variance estimators]\label{thm:consistency_variance}
    Assume that the same conditions used in Theorem \ref{thm:asymp_normal} hold. Then under $\kt, C_\alpha, \M, \lf = \bigO{1}$ and $\tnf = \bigO{n^{-1}}$,
    $$\|(\hat\sigma_j^2)^{(a,b)} - \var((\Theta_j)_{\cdot, a}^* W (\Theta_j)_{\cdot, b})\|_\infty = \bigO{d\thres + \frac{d^2}{n(2m+1)}}.$$
    Additionally under Assumption \ref{asn:8th_moment},
    $$\|(\hat\sigma_j^2)^{(a,b); \mathrm{HAC}} - \var((\Theta_j)_{\cdot, a}^* W (\Theta_j)_{\cdot, b})\|_\infty = \bigOP{d\thres + \frac{1}{2m+1} + \frac{d^2}{n(2m+1)}}.$$
    Similar bounds hold for $(\hat\delta_j^2)^{(a,b)}$ and $(\hat\delta_j^2)^{(a,b);\mathrm{HAC}}$ respectively.
\end{theorem}

The proof is deferred to Appendix \ref{pf:consistency_variance}. Assumption \ref{asn:8th_moment} can also be viewed as similar to \citet[Equation 4.3.10]{brillinger2001time}, which is required for ensuring the existence of higher-order cumulants of DFTs. Similar consistency results can alternatively be achieved by imposing standard mixing conditions.

\paragraph{Confidence ellipsoid} For a complex-valued entry of the spectral precision matrix, uncertainty is naturally represented jointly through its real and imaginary parts. We therefore work with the realified two-dimensional representation $\varphi(z) = (\re(z), \im(z))^\top\in\mathbb{R}^2, z\in \CC$. For the $(a,b)$th entry at frequency $\omega_j$, let $\hat r_j^{(a,b)} := \varphi\left((\thetahatdb_j)_{a,b}\right)$ and $r_j^{(a,b)} := \varphi\left((\Theta_j)_{a,b}\right)$. By Theorems \ref{thm:asymp_normal} and \ref{thm:consistency_variance}, $\sqrt{2m+1} \left(\hat r_j^{(a,b)} - r_j^{(a,b)}\right)$
is asymptotically normal with covariance matrix consistently estimated by
$(\hat{\Sigma}_j)^{(a,b)}$ (or $(\hat{\Sigma}_j)^{(a,b);\mathrm{HAC}}$). Hence the corresponding studentized bilinear form is
\[
    \hat D_{j,m}^{(a,b)}(r)
    :=
    \left(\hat r_j^{(a,b)} - r\right)^\top
    \left\{(\hat{\Sigma}_{j})^{(a,b)}\right\}^{-1}
    \left(\hat r_j^{(a,b)} - r\right),
    \qquad r \in \mathbb R^2 .
\]
The joint $100(1-\alpha)\%$ asymptotic confidence region for
$\varphi\left((\Theta_j)_{a,b}\right)$ is then
$$\mathcal C_{j,m}^{(a,b)}(\alpha)=\left\{r \in \mathbb R^2:(2m+1)\hat D_{j,m}^{(a,b)}(r)\leq \chi^2_2(1-\alpha)\right\},$$
where $\chi_2^2(1-\alpha)\equiv -2\log \alpha$ is the $(1-\alpha)$-quantile of a chi-squared random variable with 2 degrees of freedom. Equivalently, this gives an elliptical confidence region for the complex entry
$(\Theta_j^*)_{a,b}$ through its real and imaginary coordinates.

\subsection{Multiple hypotheses test}\label{subsec:multiple_hypotheses}

When simultaneously testing conditional associations across all coordinates of the SPM, the probability of committing at least one type I error on all entries increases exponentially with the dimension $p$. To maintain statistical power while strictly limiting the proportion of false positives among the rejected hypotheses, we employ False Discovery Rate (FDR) control.

Consider $\mathcal{Q} := \{(a,b)\in[p]^2,a < b\}$. The null hypothesis for the test at $(a,b)$\textsuperscript{th} entry is
\begin{equation}\label{eq:H0_ab}
    H_0^{(a,b)}(j): \re(\Theta_j)_{a,b}=0 \ \& \ \im(\Theta_j)_{a,b}=0.
\end{equation}
We compute the threshold $\hat{\tau}$ to control the FDR similar to \citet{liu2013gaussian}. For each $(a,b)$ node, we reject the null hypothesis $H_0^{(a,b)}(j)$ if
\begin{displaymath}
    (2m+1) \hat D_{j,m}^{(a,b)}(0) > \hat{\tau},
\end{displaymath}
where the threshold is obtained by
\begin{equation}\label{e:thres_fdr}
    \hat{\tau} 
    = \inf \left\{\tau\in(0,2\log|\mathcal{Q}|] : \widehat{\text{FDP}}_{j,m}(\tau) \leq \alpha \right\},
\end{equation}
and $\widehat{\text{FDP}}_{j,m}(\tau)$ is the estimated false
discovery proportion defined as
\begin{equation}\label{eq:fdp}
\widehat{\mathrm{FDP}}_{j,m}(\tau) :=
\frac{\exp(-\tau/2)|\mathcal{Q}|}{\max\{1,\sum_{a,b\in\mathcal{Q}}(2m+1) \hat D_{j,m}^{(a,b)}(0) \geq\tau \}},
\end{equation}
and $|\mathcal{Q}| = p(p-1)/2$. The derivation that the threshold in \eqref{e:thres_fdr} guarantees false discovery control on $\mathcal{Q}$ is omitted because it follows directly from the standard GGM literature (\citet{liu2013gaussian}; see also \citet[Section 3.2]{krampe2025frequency}, a special case for fixed frequency).

\section{Experimental results}\label{sec:numerical}

In this section, we conduct numerical studies on the deCGLASSO estimator using simulated data and the real fMRI data.

\begin{figure}[!t]
\centering
\includegraphics[width=0.85\linewidth,height=0.2\textheight]{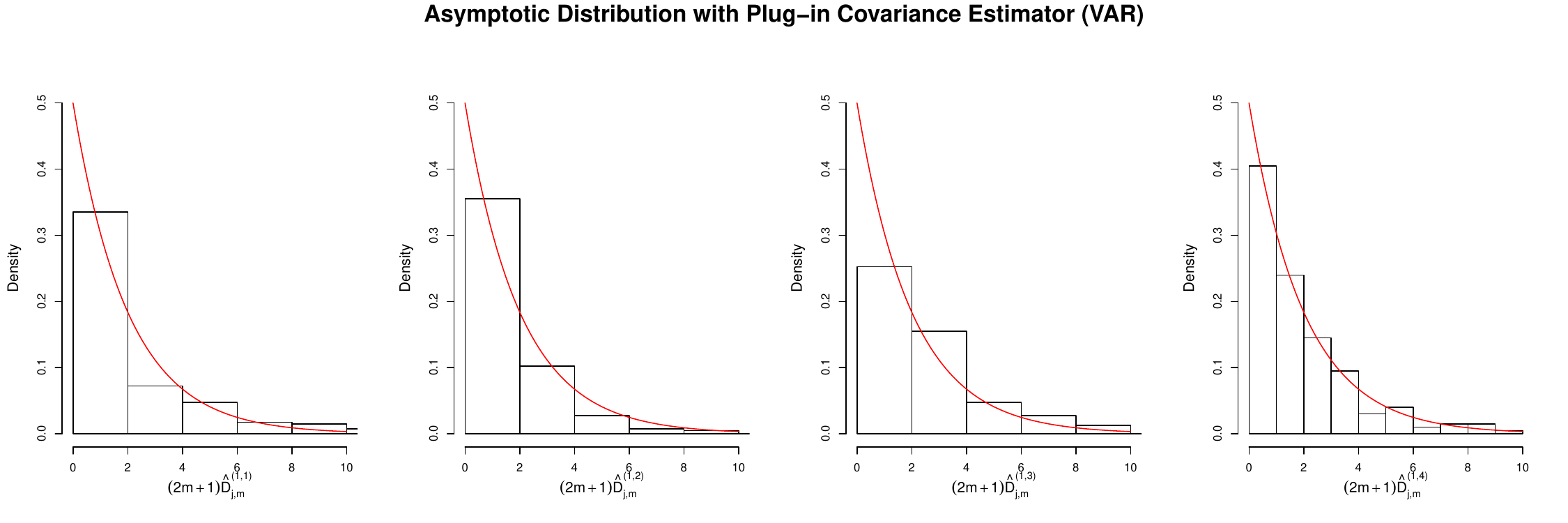}
\centering
\centering
\includegraphics[width=0.85\linewidth,height=0.2\textheight]{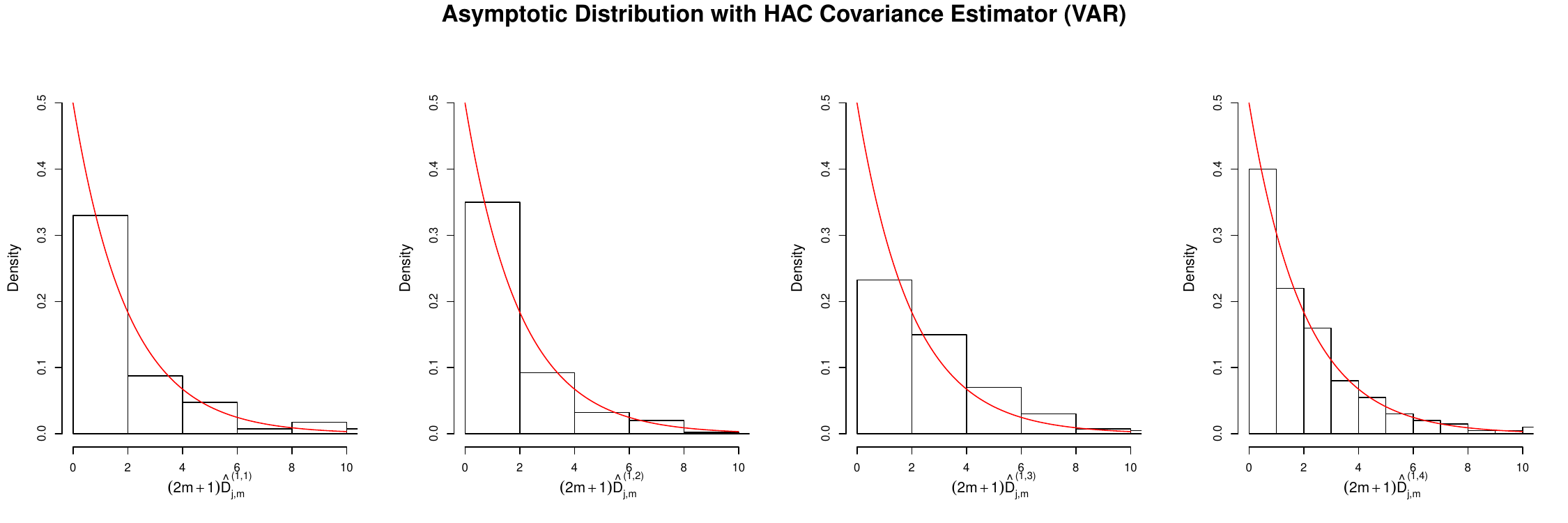}
\centering
\caption{\textbf{Histograms for $(2m+1)\hat D_{j,m}^{(a,b)}(r)$, $(a,b)\in\{(1,1),(1,2),(1,3),(1,4)\}$ at $\omega=\pi/2$.} The sample size is $n=2500$ ($m = \sqrt{n}=50$), and the number of variables is $p=100$.}
\label{fig:histogram}
\end{figure}

\subsection{Simulation studies}\label{subsec:simulation}

We demonstrate the finite-sample performance of the deCGLASSO estimator by computing empirical coverage of confidence ellipsoids, FDR, and power of multiple hypothesis tests for fixed frequency. We focus on the frequency of interest to $\omega = \pi/2$ with additional results for $\omega = \pi/100$ deferred to Appendix \ref{app:more_table}, referred to as the non-zero frequency and almost-zero frequency of interest, respectively. During the estimation, the tuning parameter is fixed as $\lambda=\sqrt{\log p/(2m+1)}$ for the deCGLASSO estimator across all setups.

\subsubsection{Empirical coverage and area of the confidence region}\label{subsubsec:coverage}

First, we empirically demonstrate that the proposed confidence regions attain the desired coverage level as the sample size increases. For $p$-dimensional vector autoregressive and moving average process of orders $q,r$ (VARMA$(q,r)$) process $X_t=\sum_{h=1}^q A_h X_{t-q} + \epsilon_{t} - \sum_{h=1}^r B_h\epsilon_{t-h}$ $\epsilon_t\sim\mathrm{WN}(0,\Sigma)$, the SPM for the model is 
\begin{equation}\label{eq:spectral_precision_VARMA}
    \Theta_j^* 
    = 2\pi (\mathcal{A}(e^{-\i w_j}))^*(\mathcal{B}^{-1}(e^{-\i w_j}))^*\Sigma^{-1}\mathcal{B}^{-1}(e^{-\i w_j})(\mathcal{A}(e^{-\i w_j})).
\end{equation}
where $\mathcal{A}(z) := I_p - \sum_{h=1}^q A_hz^h$ and $\mathcal{B}(z) := I_p - \sum_{h=1}^r B_h z^h$. Since vector autoregressive of order $q$ (VAR($q$)) and vector moving average of order $r$ (VMA($r$)) are also sub-models of VARMA$(q,r)$, all spectral precision matrices $\Theta_j^*$ for the data generating models below can be obtained by \eqref{eq:spectral_precision_VARMA}. We consider the following two data generating processes:
\begin{itemize}
    \item[(a)] \textbf{VAR(1)}: $X_t$ follows $X_t = A_1 X_{t-1} + \epsilon_t$, where $\epsilon_t \sim \nor(0, \Sigma)$ with $\Sigma = I_p$, and $A_1=\textrm{upperdiag}(0.5,-0.3)$. 
    \item[(b)] \textbf{VMA(3)}: $X_t$ follows $X_t = \varepsilon_t - B_1\varepsilon_{t-1} - B_2\varepsilon_{t-2} - B_3\varepsilon_{t-3},$ $ \varepsilon_t \overset{i.i.d.}{\sim} \normal{0}{\sigma^2 I_p}$. The moving-average coefficient matrices are $B_\ell = I_{p/5}\otimes M_\ell,\ \ell=1,2,3$. For $a,b=1,\ldots,5$, the entries of the block matrices are Toeplitz-structured, $(M_1)_{a,b}=0.9^{|a-b|+1},\
    (M_2)_{a,b}=0.6^{|a-b|+1},\ (M_3)_{a,b}=0.3^{|a-b|+1}$.

\end{itemize}

For the performance metric, we calculate the \textit{average coverage probability} over the set $A\subseteq [p]^2$ as
\begin{displaymath}
    \textrm{Avgcov}_{j}(A) := \frac{1}{|A|}\sum_{(a,b)\in A} \mathbb{P}_N\indic\left(r_j^{(a,b)} \in \mathcal{C}_{j,m}^{(a,b)}(\alpha)\right),
\end{displaymath}
We consider the edge set $E$ and its complement $E^c := [p]^2 \setminus \{E\cup\{(1,1),\ldots,(p,p)\}\}$ to compute $\textrm{Avgcov}_{j}(E)$ and $\textrm{Avgcov}_{j}(E^c)$, respectively. We also report \textit{average area} of $r_j^{(a,b)}$ using the formula
\begin{displaymath}
    \textrm{Avgarea}_{j}(A) 
    := \frac{\pi}{|A|}\cdot\frac{\chi_2^2(1-\alpha)}{2m+1}\sum_{(a,b)\in A} \sqrt{\det\left(\hat{\Sigma}_j^{(a,b)}\right)}.
\end{displaymath}
This metric is considered for the sets $E$ and $E^c$ separately to obtain $\textrm{Avgarea}_{j}(E)$ and $\textrm{Avgarea}_{j}(E^c)$. We use $\alpha=0.05$ for all setups.

We consider three methods of constructing confidence regions that are compared depending on the variance formula described in Section \ref{subsec:asymptotic_variance_covariance}, i.e., with population asymptotic covariance matrix (Pop), the plug-in estimators (Plug-in), and the HAC (HAC) estimators.

\begin{table}[!t]
\centering
\renewcommand{\arraystretch}{1.15} %
\resizebox{0.8\columnwidth}{!}{%
\begin{tabular}{cccccc}
\toprule
$n$ & Method 
& $\textrm{Avgcov}_{j}(E)$ 
& $\textrm{Avgarea}_{j}(E)$ 
& $\textrm{Avgcov}_{j}(E^c)$ 
& $\textrm{Avgarea}_{j}(E^c)$ \\
\midrule
\multirow{3}{*}{400} 
& Pop     & 0.982 (0.009) & 21.690 (0.000) & 0.978 (0.003) & 16.250 (0.000) \\
& Plug-in & 0.930 (0.018) & 18.711 (0.948) & 0.964 (0.004) & 13.608 (0.668) \\
& HAC     & 0.904 (0.021) & 15.880 (0.670) & 0.935 (0.004) & 11.585 (0.387) \\
\cline{1-2}
\multirow{3}{*}{900}
& Pop     & 0.977 (0.010) & 14.578 (0.000) & 0.978 (0.003) & 10.922 (0.000) \\
& Plug-in & 0.916 (0.018) & 11.823 (0.417) & 0.957 (0.003) & 8.698 (0.300) \\
& HAC     & 0.901 (0.021) & 10.762 (0.316) & 0.939 (0.004) & 7.950 (0.209) \\
\cline{1-2}
\multirow{3}{*}{1600}
& Pop     & 0.970 (0.012) & 10.979 (0.000) & 0.978 (0.003) & 8.225 (0.000) \\
& Plug-in & 0.906 (0.020) & 8.709 (0.267) & 0.954 (0.004) & 6.444 (0.194) \\
& HAC     & 0.896 (0.021) & 8.172 (0.214) & 0.941 (0.004) & 6.054 (0.140) \\
\cline{1-2}
\multirow{3}{*}{2500}
& Pop     & 0.962 (0.012) & 8.805 (0.000) & 0.978 (0.003) & 6.596 (0.000) \\
& Plug-in & 0.896 (0.019) & 6.941 (0.195) & 0.953 (0.004) & 5.150 (0.143) \\
& HAC     & 0.887 (0.021) & 6.611 (0.171) & 0.943 (0.004) & 4.906 (0.105) \\
\bottomrule
\end{tabular}%
}
\caption{\textbf{Average coverage probabilities and coverage areas over $E$ and $E^c$ for the VAR(1) model at $\omega=\pi/2$, with $p=100$.} The reported values are averages over 200 repetitions (standard deviations in parentheses).}
\label{tab:sim_var1_nonzero_main}
\end{table}

\paragraph{Results}
Figure \ref{fig:histogram} shows the empirical distributions of $(2m+1)\hat D_{\lfloor n/4 \rfloor,m}^{(a,b)}(r)$, which are close to $\chi^2_2$ distribution for representative zero and nonzero entries of $\Theta(\pi/2)$. This suggests that the estimated covariance mainly affects the size of the confidence regions, rather than the shape of the limiting approximation. Table \ref{tab:sim_var1_nonzero_main} reports the corresponding average coverage and area over $E$ and $E^c$ for $p = 100$. The confidence regions based on the population covariance formula serve as an oracle benchmark and are marginally conservative. The feasible plug-in and HAC covariance estimators perform comparably and yield coverage close to the nominal level, with mild undercoverage mainly for entries in $E$. Across all variance estimators, the average areas decrease as $n$ increases, while the corresponding coverage probabilities remain stable. In contrast, coverage over $E^c$ is consistently close to the nominal level across the DGPs, frequencies, and support sets considered. The remaining undercoverage over $E$ is consistent with the smaller estimated areas of the feasible confidence regions and is partly reduced as $n$ increases in some settings. Additional coverage results for the VAR(1) model at $\omega=\pi/2$ and $p=50$ and 150, VAR(1) results at the near-zero frequency $\omega=\pi/100$, and results for the VMA(3) model at $\omega=\pi/2$ are reported in Tables \ref{tab:sim_var1_nonzero_appendix}, \ref{tab:sim_var1_almostzero_appendix} and \ref{tab:sim_vma3_coverage_area}, respectively.

\subsubsection{FDR and power diagnostics}\label{subsubsec:fdr_power}

Next, we perform the tests of nullity across off-diagonal entries of $\Theta(\omega)$ to compare with the PSC-based benchmark method using nodewise regression with a plug-in SPM estimator \citep{krampe2025frequency}, referred to as NWR-PSC. To ensure a fair comparison, we set the threshold of the pivotal test statistic in the benchmark method to 0. The rest of the setup remains the same.

We define the set of discoveries as $\hat{E}=\{(a,b)\in\mathcal{Q}:(2m+1) \hat D_{j,m}^{(a,b)}(0) \geq \hat{\tau}\}$ using $\hat \tau$ in \eqref{e:thres_fdr}. We consider the sets of nodes in the upper triangular regions of $E$ and report the empirical false discovery rate (FDR), the Monte Carlo average of the empirical FDP, and the empirical power respectively, where
\begin{align*}
    \mathrm{FDP} = & \frac{\sum_{a,b \in E^c}\indic\{(2m+1)\hat D_j^{(a,b)}(0) \ge \hat \tau\}}{\max\{\sum_{a,b \in \mathcal{Q}}\indic\{(2m+1)\hat D_j^{(a,b)}(0) \ge \hat \tau\}, 1\}}\\
    \mathrm{Empirical\ power} = & \frac{\sum_{a,b \in E}\indic\{(2m+1)\hat D_j^{(a,b)}(0) \ge \hat \tau\}}{|E|}.
\end{align*}
The bandwidth is selected using an extension of the GCV-based deviance criterion of \citet{ombao2001simple}, averaged over the diagonal entries of the periodogram, and is denoted by $\hat m\in \argmin_{m\in\mathcal{G}} \mathrm{GCV}(m)$ over the grid set $\mathcal G=\{1+10\ell:\ell\ge 0,\ 1+10\ell\le \lfloor n/3\rfloor\}$, where
\begin{equation}\label{eq:gcv}
    \mathrm{GCV}_{\Theta}(m) = \dfrac{\dfrac{1}{pn/2} \sum_{j=0}^{n/2} q_j \sum_{a = 1}^p \left\{-\log \det \left(\dfrac{(d_j d_j^*)_{a,a}}{(\hat f_j)_{a,a}}\right) + \dfrac{(d_j d_j^*)_{a,a} - (\hat f_j)_{a,a}}{(\hat f_j)_{a,a}} \right\}}{\left(\dfrac{2m}{2m+1}\right)^2},
\end{equation}
with $q_j = 1 - 0.5\indic\{j = 0 \text{ or } j = n/2\}$. For the deCGLASSO estimator, $\lambda$ is selected using EBIC \citep{foygel2010extended} and the experiments are conducted across 50 Monte Carlo replicates. The empirical FDR is computed as a Monte Carlo average of the FDP across the trials.

\paragraph{Results} The empirical FDR and power are reported in Table \ref{tab:fdr_power_decglasso_benchmark}. For both the VAR(1) and VMA(3) DGPs, deCGLASSO achieves higher power than NWR-PSC when using either the plug-in or the HAC variance estimator. The power of deCGLASSO increases with the sample size $n$. In contrast, NWR-PSC exhibits substantially lower power, especially for smaller values of $n$, suggesting that it requires larger sample sizes to manifest its asymptotic power behavior. While the FDR of NWR-PSC remains below the nominal level of $\alpha=0.05$ across all considered settings, the FDR of deCGLASSO marginally exceeds $0.05$ for some experiments. This exceedance of FDR diminishes as $n$ increases, during which deCGLASSO gains substantially higher power. Finally, the plug-in and HAC variance estimators show broadly similar performance: the plug-in estimator performs marginally better in the VMA(3) setting, whereas the HAC estimator performs marginally better in the VAR(1) setting.

\begin{table}[!t]
\centering
\resizebox{\linewidth}{!}{
\begin{tabular}{llcccccccc}
\toprule
& & & \multicolumn{2}{c}{deCGLASSO-plugin} 
& \multicolumn{2}{c}{deCGLASSO-HAC} 
& \multicolumn{2}{c}{NWR-PSC} \\
\cmidrule(lr){4-5} \cmidrule(lr){6-7} \cmidrule(lr){8-9}
DGP & $p$ & $n$ & FDR & Power & FDR & Power & FDR & Power \\
\midrule
\multirow{6}{*}{VAR(1)}
& \multirow{3}{*}{50}  
& 1000 & 0.029 (0.022) & 0.713 (0.079) & 0.074 (0.026) & 0.797 (0.077) & 0.000 (0.000) & 0.009 (0.010) \\
& & 2000 & 0.039 (0.021) & 0.938 (0.045) & 0.067 (0.024) & 0.953 (0.043) & 0.000 (0.000) & 0.160 (0.044) \\
& & 4000 & 0.035 (0.020) & 0.998 (0.006) & 0.055 (0.020) & 0.999 (0.005) & 0.002 (0.005) & 0.815 (0.045) \\
\cline{2-3}
& \multirow{3}{*}{100} 
& 1000 & 0.033 (0.015) & 0.638 (0.071) & 0.082 (0.018) & 0.739 (0.080) & 0.005 (0.035) & 0.013 (0.007) \\
& & 2000 & 0.037 (0.014) & 0.925 (0.028) & 0.074 (0.016) & 0.946 (0.024) & 0.001 (0.005) & 0.165 (0.156) \\
& & 4000 & 0.038 (0.015) & 0.997 (0.005) & 0.064 (0.016) & 0.998 (0.004) & 0.003 (0.004) & 0.892 (0.023) \\
\cline{1-3}
\multirow{6}{*}{VMA(3)}
& \multirow{3}{*}{50}  
& 1000 & 0.078 (0.041) & 0.854 (0.053) & 0.053 (0.022) & 0.838 (0.047) & 0.013 (0.014) & 0.558 (0.039) \\
& & 2000 & 0.075 (0.027) & 0.897 (0.041) & 0.054 (0.021) & 0.887 (0.036) & 0.027 (0.019) & 0.729 (0.029) \\
& & 4000 & 0.061 (0.025) & 0.941 (0.030) & 0.046 (0.021) & 0.932 (0.027) & 0.030 (0.018) & 0.886 (0.025) \\
\cline{2-3}
& \multirow{3}{*}{100} 
& 1000 & 0.090 (0.034) & 0.863 (0.038) & 0.059 (0.020) & 0.846 (0.033) & 0.012 (0.010) & 0.528 (0.022) \\
& & 2000 & 0.084 (0.023) & 0.902 (0.030) & 0.059 (0.012) & 0.887 (0.027) & 0.027 (0.014) & 0.697 (0.022) \\
& & 4000 & 0.071 (0.025) & 0.941 (0.022) & 0.050 (0.018) & 0.934 (0.022) & 0.036 (0.017) & 0.862 (0.016) \\
\bottomrule
\end{tabular}}
\caption{\textbf{Comparison of empirical FDR and power for deCGLASSO with plug-in and HAC variance estimators, and the benchmark method NWR-PSC \citep{krampe2025frequency}.} Entries are reported as means over 50 replicates (standard error in parentheses).}
\label{tab:fdr_power_decglasso_benchmark}
\end{table}

\begin{figure}[t]
\centering
\includegraphics[width=0.4\textwidth, trim={0.5in 0.1in 0in 0.25in}]{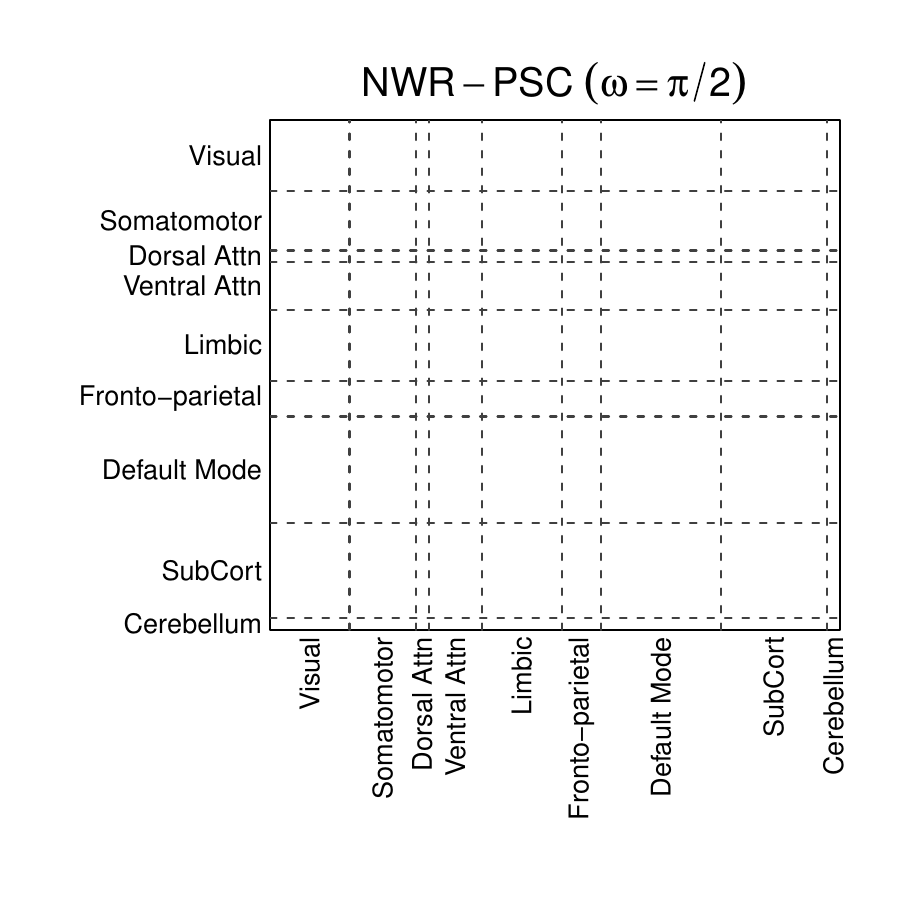}
\includegraphics[width=0.4\textwidth, trim={0.45in 0.3in 0.25in 0.25in}]{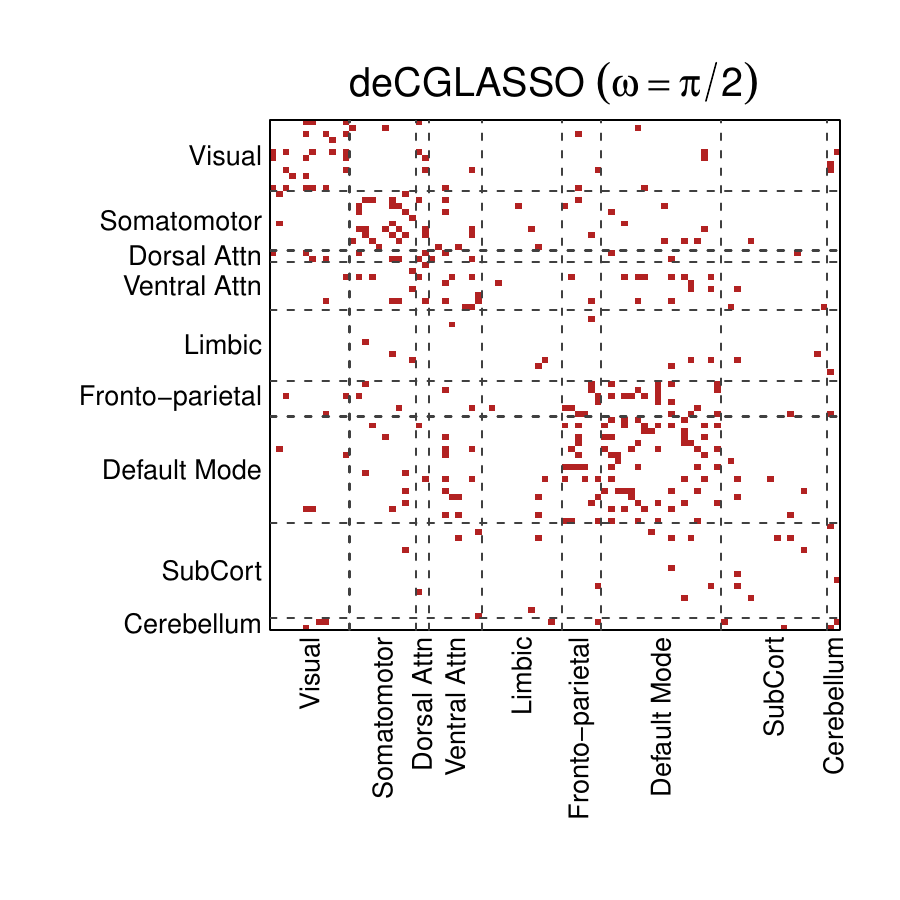}\\

\includegraphics[width=0.4\textwidth, trim={0.5in 0.3in 0.25in 0.25in}]{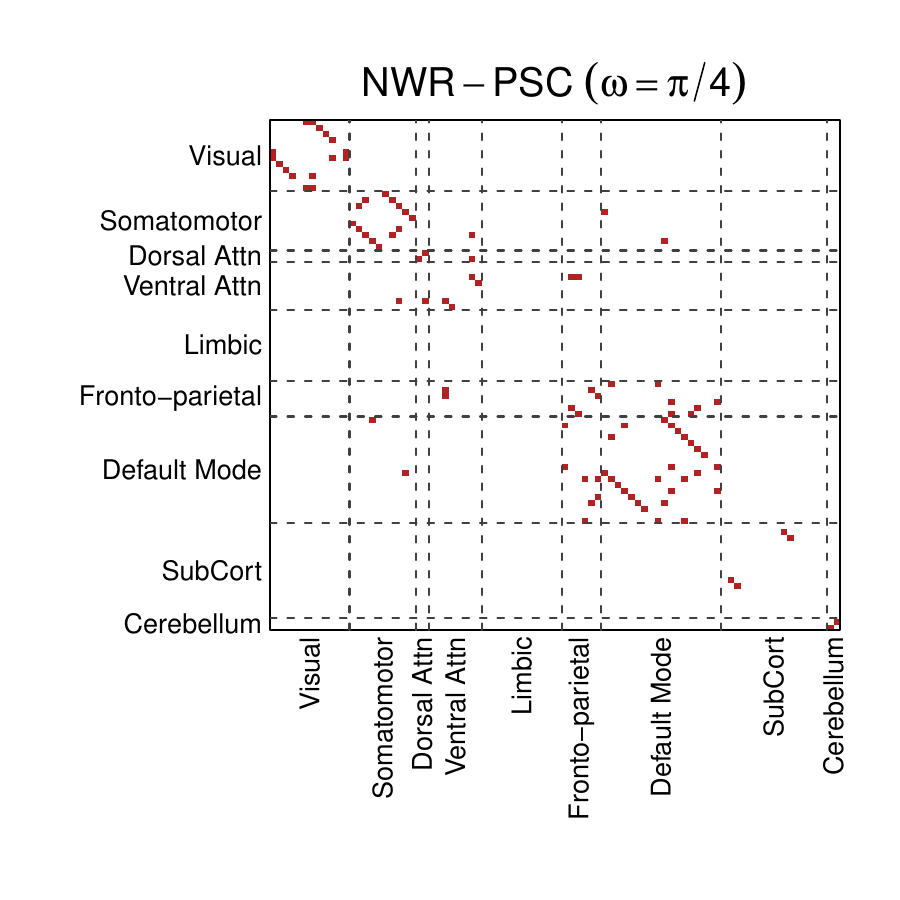}
\includegraphics[width=0.4\textwidth, trim={0.5in 0.3in 0.25in 0.25in}]{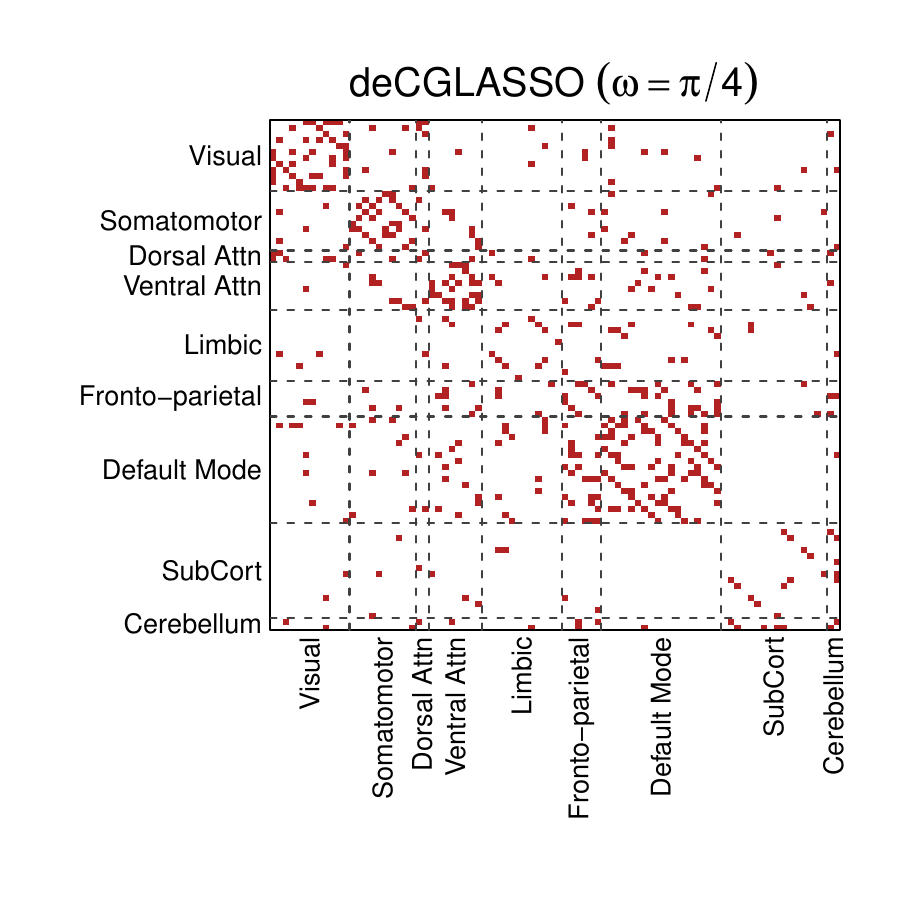}\\

\includegraphics[width=0.4\textwidth, trim={0.5in 0.3in 0.25in 0.25in}]{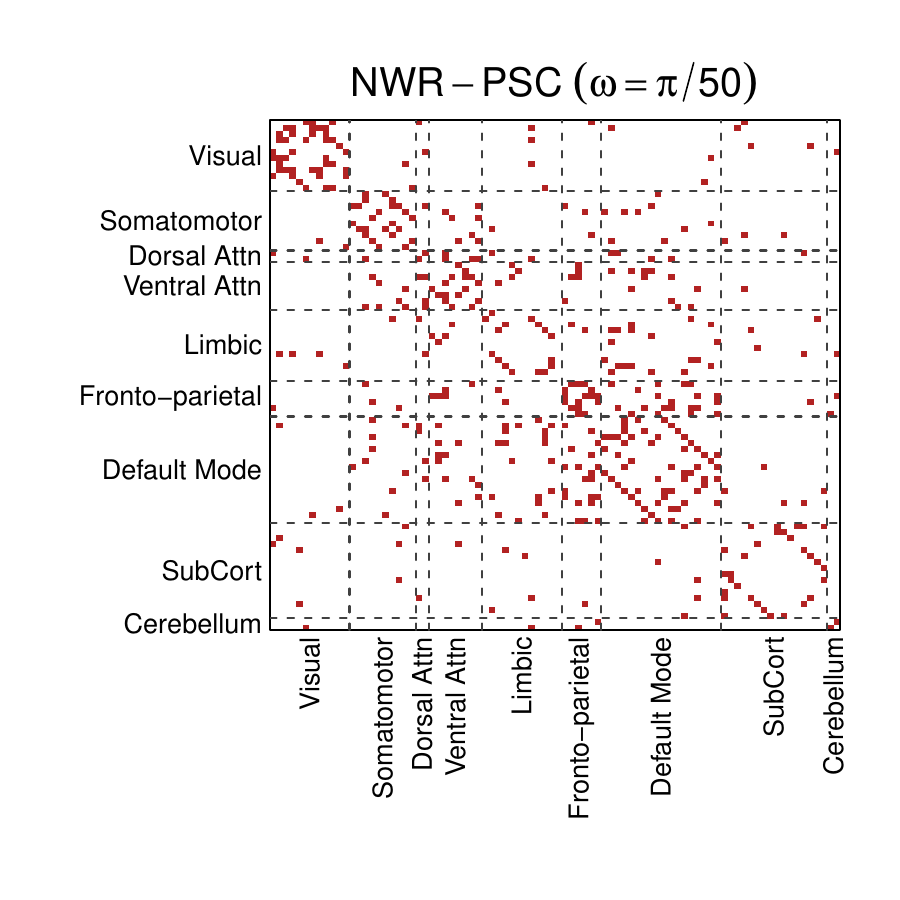}
\includegraphics[width=0.4\textwidth, trim={0.5in 0.3in 0.25in 0.25in}]{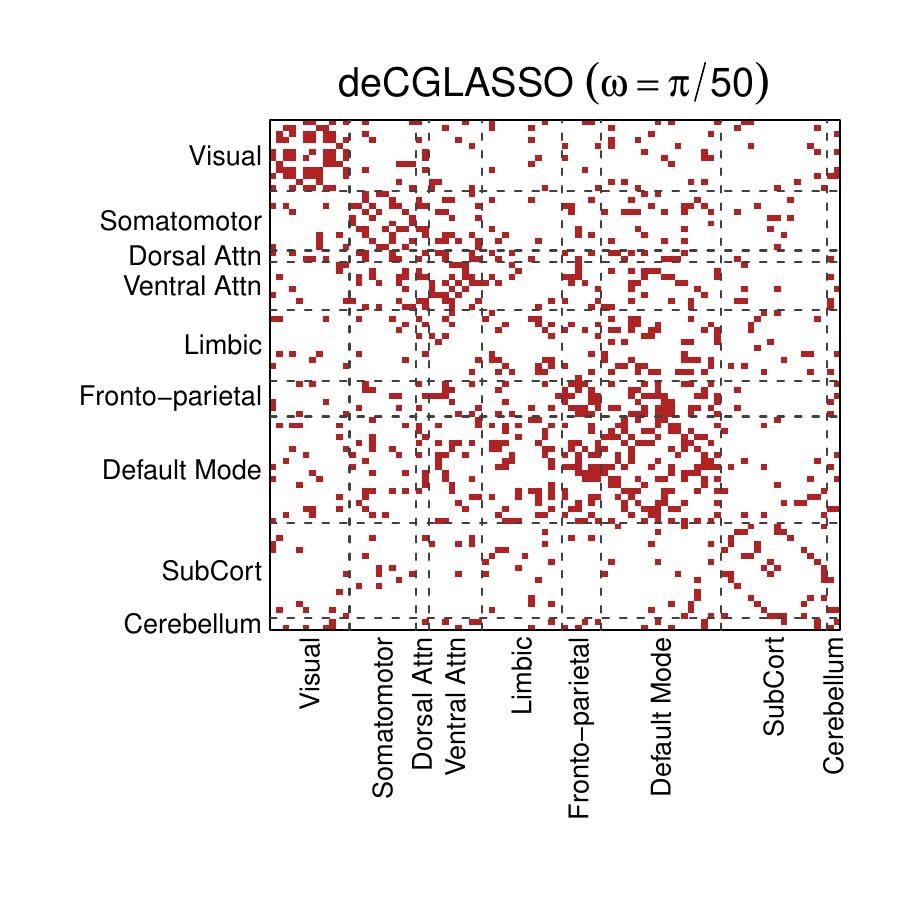}
\caption{\textbf{Heatmaps of significant edge adjacency matrices for one representative subject using the benchmark NWR-PSC and our method deCGLASSO at $\omega=\pi/2,\pi/4,$ and $\pi/50$.} Red entries indicate the significant entries selected by the FDR thresholding procedure; white entries indicate non-significant entries.}
\label{fig:heatmap}
\end{figure}

\subsection{fMRI data analysis}\label{subsec:data}

We apply our inference procedure to the resting-state fMRI dataset from the Human Connectome Project (HCP) -- Young Adult S1200 release \citep{van2013wu}. The goal is to check how deCGLASSO performs in extracting the FC matrices and how that varies across frequencies as compared to the benchmark NWR-PSC \citep{krampe2025frequency}.

We use parcellated gray matter of the $p=86$ brain regions (cortical), which are further mapped to one of the nine functional groups (visual, somatomotor, dorsal attention, ventral attention, limbic, default mode, fronto-parietal, subcortical, and cerebellar networks) by Yeo's functional parcellation \cite{yeo2011organization}. Each scan consisted of $n=1200$ time points, with four scans per subject. We choose $m$ with the GCV criterion in \eqref{eq:gcv} and consider the frequency-specific FC at $\omega = \pi/2, \pi/4$ and $\pi/50$. Due to limited number of DFTs within the bandwidth relative to the number of brain regions, the averaged periodogram is not always invertible. Hence we first compute the averaged periodogram estimator for each scan and then further average them to obtain an invertible $\hat f(\omega)$, and use it as an input to deCGLASSO and NWR-PSC by replacing the averaged periodogram appearing in \citet[Equation 12]{krampe2025frequency} and their corresponding SPM solver by the average of the averaged periodograms across the four scans.

We implement the multiple testing procedure in Section \ref{subsec:multiple_hypotheses} for selecting the significant entries. For each edge $(a,b)$ in the $\omega$-specific graph of one subject, we define the edge-specific q-value \citep{storey2003statistical} by
$$q_{j,m}^{(a,b)} = \inf_{0\le \tau  \le \min\left\{(2m+1)\hat D_{j,m}^{(a,b)}(0),\, 2\log |\mathcal{Q}|\right\}}
\widehat{\mathrm{FDP}}_{j,m}(\tau).$$
where $\widehat{\text{FDP}}_{j,m}(\tau)$ is defined in \eqref{eq:fdp}. Thus, $q_{j,m}^{(a,b)}$ is the smallest estimated FDP level at which edge $(a,b)$ would be selected by the thresholding rule. We use volcano plots to display each edge through the pair
$(\sqrt{|(\widetilde\rho(\omega))_{a,b}|}, -\log_{10} q_{j,m}^{(a,b)}).$
The horizontal coordinate measures the magnitude of the debiased PSC estimator capturing the signal strength, and the vertical axis measures FDR-adjusted evidence.

\paragraph{Results} 

Figure~\ref{fig:heatmap} shows heatmaps of significant edges across selected frequencies for NWR-PSC and deCGLASSO for a single user. Overall, deCGLASSO selects more significant entries than NWR-PSC at the higher frequencies considered. At $\omega=\pi/2$, the NWR-PSC heatmap is empty, whereas deCGLASSO selects several edges, with visible concentrations within and across blocks corresponding to large-scale functional systems, including the visual, somatomotor, fronto-parietal, and default-mode networks. At $\omega=\pi/4$, NWR-PSC begins to select some entries in these blocks, while deCGLASSO yields a denser pattern of selected entries, including in the subcortical block. Several selected regions also appear consistent with bilateral homologue structure between right and left hemispheric components, as also observed in \citet{deb2024regularized}. At the lower frequency $\omega=\pi/50$, both methods select substantially more entries, with deCGLASSO showing a denser pattern near these homologue regions. This qualitative pattern is consistent with the well-known observation that resting-state BOLD functional connectivity is largely driven by low-frequency fluctuations \citep{biswal1995functional}.

\begin{figure}[!t]
\centering
\includegraphics[width=0.4\textwidth]{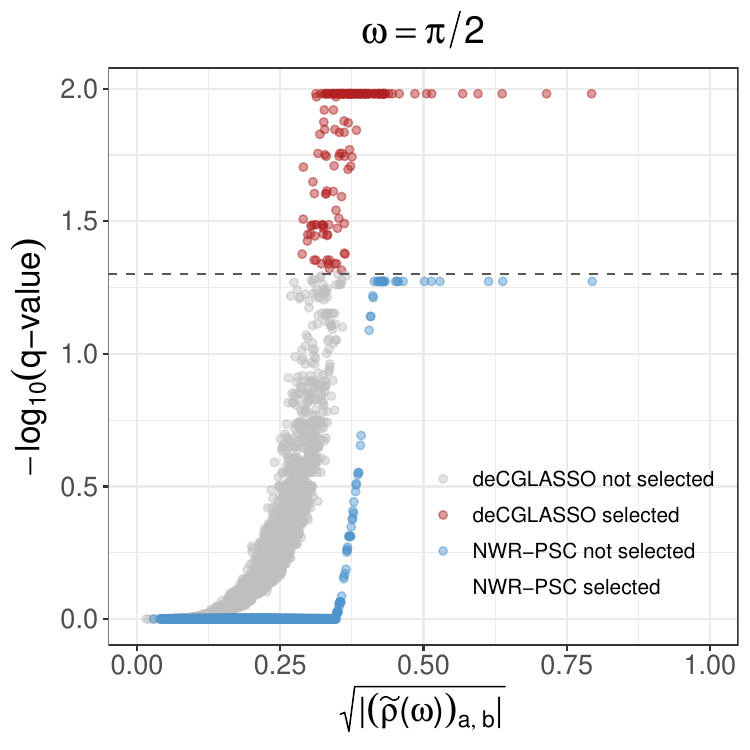}\hspace{10pt}
\includegraphics[width=0.4\textwidth]{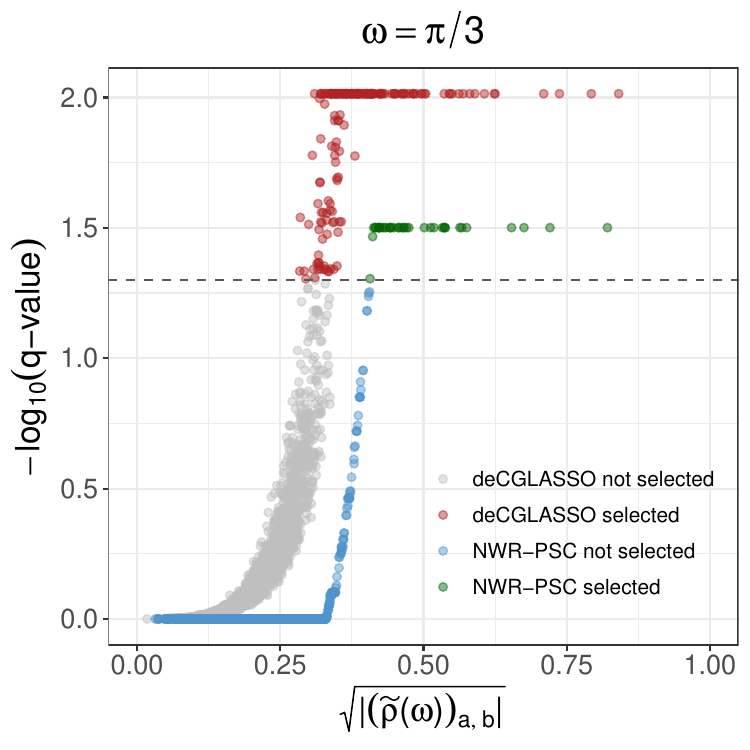}\\
\vspace{20pt}

\includegraphics[width=0.4\textwidth]{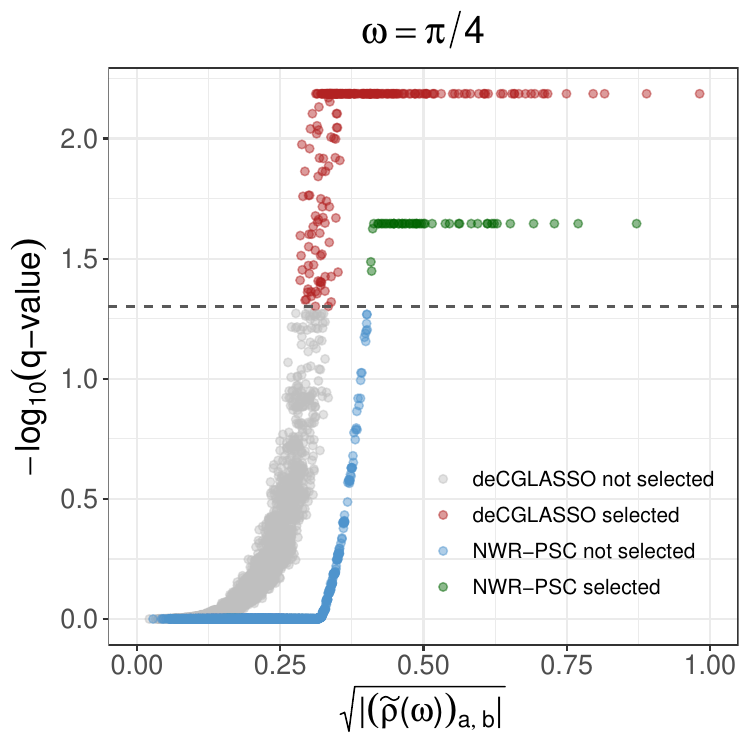}
\includegraphics[width=0.4\textwidth]{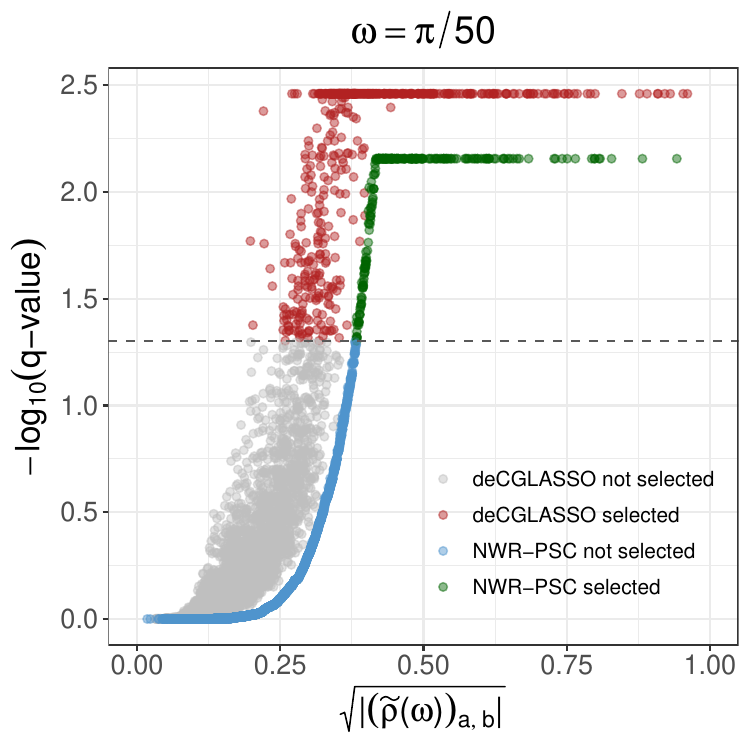}
\caption{\textbf{Volcano plots of deCGLASSO and NWR-PSC at $\omega=\pi/2, \pi/3, \pi/4, \pi/50$.} Each point is a candidate edge $(a,b)$, with $x$-axis $\sqrt{ (|\widetilde\rho (\omega))_{a,b}|}$ and $y$-axis $ -\log_{10} (\text{q-value})$ for $(a,b)$. The dashed line marks the threshold $-\log_{10}(0.05)\approx 1.301$ corresponding to the FDR level $\alpha=0.05$.}
\label{fig:volcano}
\end{figure}

Across the frequencies $\omega = \pi/2, \pi/3, \pi/4$ and $\pi/50$, deCGLASSO selects more significant edges than NWR-PSC, as shown in the volcano plot Figure~\ref{fig:volcano}. At $\omega=\pi/2$, deCGLASSO selects a substantial number of entries whereas NWR-PSC selects no entries, consistent with their corresponding heatmaps in Figure~\ref{fig:heatmap}. At $\omega=\pi/3$ and $\omega=\pi/4$, NWR-PSC starts to select more significant edges, but its points tend to lie to the right and below those from deCGLASSO -- indicating that NWR-PSC tends to select entries with larger magnitudes of the square-rooted debiased PSC estimator $\sqrt{|(\widetilde \rho(\omega))_{a,b}|}$, while their evidence relative to the FDR threshold remains smaller than that of the entries selected by deCGLASSO. In contrast, deCGLASSO selects a broader set of entries, including many with moderate partial coherence values but stronger threshold-relative evidence. At the lower frequency $\omega=\pi/50$, both methods produce more discoveries, with selected entries appearing farther above the threshold; however, the separation between the two volcano plots persists, and deCGLASSO continues to yield a denser set of significant edges. At $\omega=\pi/50$, NWR-PSC exhibits a smooth evidence pattern, whereas deCGLASSO shows larger dispersion among moderate-strength edges -- capturing the entry-specific uncertainty in the debiased precision-scale. Similar patterns are observed for another example user, as in Figures~\ref{fig:heatmap2} and~\ref{fig:volcano2}. These results suggest that deCGLASSO detects a larger set of frequency-specific conditional associations, while NWR-PSC behaves more conservatively by selecting only entries with comparatively larger estimated partial coherence magnitudes.

\section{Discussion}\label{sec:discussion}

In this work, we develop an inference framework for SPMs in high dimensions. Inference in this setting is fundamentally more challenging than its i.i.d. counterpart due to the lack of asymptotic distribution and variance expressions in high dimensions, the complex-valued nature of the debiased estimators, and the subject to additional biases from finite-sample truncation and smoothing. We proposed the deCGLASSO estimator and established its entry-wise asymptotic normality via a CLT for bilinear forms of complex-valued multivariate time series. Key contributions are the derivation of the asymptotic covariance matrix of the entries of the deCGLASSO estimator, and control of regularization bias along with truncation and smoothing biases for valid inference. The framework is validated in terms of coverage, FDR, and power performance through simulation studies and an application to real fMRI data.

There are several possible directions for future research. First, our framework could be extended to multiple testing problems over arbitrary combinations of frequencies. This would allow more flexible inference on spectral connectivity patterns and provide a step toward full graphical modeling for time series. Such an extension is also motivated by the fact that, in many applications, only a small number of frequencies carry the most relevant signal; targeting these frequencies may therefore lead to improved statistical power. Second, the bandwidth selection procedure in this work is motivated by a GCV criterion \citep{ombao2001simple}. This criterion could be further improved, both theoretically and computationally, by incorporating the full Whittle likelihood of the nearby DFTs. These two directions are closely related, since both require efficient estimation of the spectral precision matrix across neighboring frequencies. A promising computational direction is to develop online or sequential algorithms that exploit the shared DFTs appearing in the local frequency neighborhoods of nearby target frequencies. Finally, it would be of interest to extend the proposed methodology to non-stationary settings \citep{basu2023graphical}.

\section*{Acknowledgements}

SB acknowledges partial support from NSF CAREER award DMS-2239102, NSF awards DMS-1812128, DMS-2210675, and NIH awards R01GM135926, R21NS120227. The authors are grateful to Prof. Suhasini Subba Rao for valuable input in the theoretical foundations of the work and insightful discussions, and to Prof. Amy Kuceyeski for sharing the fMRI data sets.

{\small
\bibliographystyle{abbrvnat}
\bibliography{00-0-all}
}

\appendix

\etocdepthtag.toc{appendix}
\begin{center}\bfseries\large Table of Contents for the Appendix\end{center}
\etocsettocstyle{}{}
{\etocsettagdepth{appendix}{subsection}
\etocsettagdepth{main}{none}
\tableofcontents}

\setcounter{section}{0}
\setcounter{figure}{0}
\setcounter{table}{0}
\setcounter{equation}{0}
\renewcommand{\thesection}{A\arabic{section}}
\renewcommand{\thepage}{\arabic{page}}
\renewcommand{\thetable}{A\arabic{table}}
\renewcommand{\thefigure}{A\arabic{figure}}

\section{Proof of Proposition \ref{prop:reg_bias}}\label{pf:reg_bias}

We use Lemma \ref{lem:remainder_conditional} with the following bound:
\begin{align*}
    \|\Theta_j W\|_\infty = & \max_{a\in [p]}|(\Theta_j)_{\cdot,a}^* W \e_b |\\
    = & \max_{a\in [p]} \left\| (\Theta_j)_{\cdot,a} \right\|_2 \left| \left( \frac{(\Theta_j)_{\cdot,a}}{\|(\Theta_j)_{\cdot,a}\|_2} \right)^* W \e_b \right|\\
    \le & \M \thres \quad \text{with probability}\ \ge 1 - c\exp(-c' \log p),
\end{align*}
where the inequality holds due to Lemma \ref{lem:single_deviation}. Hence using the conditions in the statement, we have
\begin{align*}
    \|R_{n,p}(j)\|_\infty 
    = & \bigO{ \frac{1}{\gamma} \kt \max\left\{ d\thres^2\M, \frac{1}{\gamma} \kt d^2 \thres^3, \frac{1}{\gamma} \kt \kf d\thres^2 \right\}}\\
    = & \bigO{ \frac{1}{\gamma^2} \kt^2\kf d\thres^2 },
\end{align*}
where the last equality holds since $d \thres \leq [6 \kt^2 \kf^3 C_\gamma]^{-1}$ and $\gamma\in(0,1)$, $\kf,\kt\geq1$. Then $\kf\leq\M$ completes the proof. \pfend

\begin{lemma}[Element-wise deviation bound]\label{lem:single_deviation}
Let $\{X_t\}_{t\in[n]}$ be observations from a stable Gaussian centered time series satisfying Assumption \ref{asn:summable}. Then the bias of $\hat f_j$ is controlled as
\begin{equation}\label{eq:avg_periodogram_bias}
    \| \EE[\hat f_j] - f_j \|_\infty \le \vmnf.
\end{equation}
Additionally for $m \succsim\M^2 \log p$ and $m \le n/(4\M\lf)$ and $A>0$, there exist constants $c, c' > 0$ such that
\begin{equation}\label{eq:single_deviation_max}
    \PP\left(\big\| W \big\|_\infty \ge \thres\right) \le c' \exp(-c\log p)
\end{equation}
\end{lemma}
A proof of Lemma \ref{lem:single_deviation} is in \citet[Lemma B.4]{deb2024regularized}. \pfend

\begin{lemma}[Conditional control on the remainder]\label{lem:remainder_conditional}
    Under the conditions of Theorem \ref{prop:reg_bias} and on the set $\{\| W \| < \thres \}$,
    $$ \|R_{n,p}(j)\|_\infty = \bigO{ \frac{1}{\gamma} \kt \max\left\{d\thres \|\Theta_j W_j\|_\infty, \frac{1}{\gamma} \kt d^2 \thres^3, \frac{1}{\gamma} \kt \kf d\thres^2 \right\} }. $$
\end{lemma}
Lemma \ref{lem:remainder_conditional} follows the same result as Bound I in \citet[Lemma 3]{jankova2015confidence}, with replacement of estimated and population covariance matrices with $\hat f_j$ and $f_j$, respectively. \pfend

\section{Proof of Theorem \ref{thm:asymp_normal}}\label{pf:asymp_normal}

From \eqref{eq:bias_variance_decomposition}, we have
\begin{align}
    & -\sqrt{2m+1}\left((\thetahatdb_j)_{a,b} - (\Theta_j)_{a,b}\right) \nonumber\\
    &= \sqrt{2m+1}\ \e_a^\top \left(\Theta_j \hat f_j \Theta_j - \Theta_j\right) \e_b + R_{n,p}. \label{eq:debias_expansion}
\end{align}
As long as $d\thres ^2 = o(\gamma^2/\M \kt^2)$, by the result of Theorem \ref{prop:reg_bias}, $\|R_{n,p}\|_\infty = o(1)$. The first term in \eqref{eq:debias_expansion} can be written as
\begin{equation}\label{eq:decomp}
    \sqrt{2m+1}\ (\Theta_j)_{a,\cdot} (\hat f_j - \EE[\hat f_j]) (\Theta_j)_{\cdot,b} + \sqrt{2m+1}(\Theta_j)_{a,\cdot} (\EE[\hat f_j] - f_j) (\Theta_j)_{\cdot,b}.
\end{equation}
Since $\sqrt{m}\M^2 \vmnf = o(1)$, the second term in \eqref{eq:decomp} is $o(1)$. Now we derive the asymptotic distribution of the first term in \eqref{eq:decomp}. Note that 
\begin{align}
    & (\Theta_j)_{a,\cdot} \hat f_j (\Theta_j)_{\cdot,b} \nonumber \\
    = & (\Theta_j)_{a,\cdot} \left( \frac{1}{2m+1} \sum_{k \in \wmj} \left[ \frac{1}{\sqrt{2\pi n}} \sum_{t = 1}^n X_t e^{-\i t\omega_k} \right] \left[ \frac{1}{\sqrt{2\pi n}} \sum_{t = 1}^n X_t^\top e^{\i t\omega_k} \right]\right) (\Theta_j)_{\cdot,b} \nonumber \\
    = & \frac{1}{2\pi} \sum_{1\le t, t'\le n} \left(\frac{1}{n} \cdot \frac{1}{2m+1}\sum_{k \in \wmj} e^{-\i (t-t')\omega_k}\right)\left((\Theta_j)_{a,\cdot} X_t\right) \left((\Theta_j)_{b,\cdot} X_{t'}\right)^{\top}.\label{eq:bilinear_form_simplified}
\end{align}
Define 
\begin{equation}\label{eq:dirichlet_kernel}
    D_m(x) := \sum_{h = -m}^m e^{-\i h x} = 1 + 2\sum_{h=1}^m \cos(hx),
\end{equation}
referred as the \emph{Dirichlet kernel}. Thus, 
\begin{align*}
    (\Theta_j)_{a,\cdot} \hat f_j (\Theta_j)_{\cdot,b} 
    = \sum_{1\le t,t' \le n}\left( \frac{D_m(2\pi (t-t')/n)}{2\pi n(2m+1)} e^{-\i (t-t')\omega_j}  \right)\left((\Theta_j)_{a,\cdot} X_t\right) \left((\Theta_j)_{b,\cdot} X_{t'}\right)^{\top}
\end{align*}

Finally, we use Lemma \ref{lem:cmplx_clt} to obtain that the asymptotic distribution of $(\Theta_j)_{a,\cdot} \hat f_j (\Theta_j)_{\cdot,b}$ is as in \eqref{eq:cmplx_clt}, where the asymptotic variances and the covariance are given in Theorem \ref{thm:asymp_normal}. Hence, the proof is done. \pfend

Now we state the necessary lemmas for Theorem \ref{thm:asymp_normal}. For notational convenience, we first introduce the notation adapted to our setting and then briefly rederive the joint CLT for multiple time series.

\paragraph{Physical dependence measures} For a $p$-dimensional causal process $\{X_t\}_{t\in\ZZ}$ represented as
$$ X_t = G(\ldots, \varepsilon_{j-1}, \varepsilon_j, \varepsilon_{j+1}, \ldots, \varepsilon_t), \quad j \le t, $$
a coupled process, by replacing the $j$\textsuperscript{th} noise with an i.i.d. copy, is denoted by
$$ X_{t|\{j\}} = G(\ldots, \varepsilon_{j-1}, \widetilde\varepsilon_j, \varepsilon_{j+1}, \ldots, \varepsilon_t), \quad j \le t, $$
where $\varepsilon_j$ and $\widetilde\varepsilon_j$ are i.i.d. The functional dependence \citep{wu2018asymptotic} measure is defined as
$$ \delta_{t,q}^{(a)} := \|X_t^{(a)} - X_{t|\{0\}}^{(a)}\|_q, $$
where for any process $\{Z_t\}_{t\in\ZZ}$, its $\ell_q$-norm is denoted by $\|Z_t\|_q := \left(\sum_{a = 1}^p \EE[|Z_t^{(a)}|^q]\right)^{1/q}$ and $\|Z_t\| \equiv \|Z_t\|_2 $. $\delta_{t,q}^{(a)}$ determines the dependence of $X_t^{(a)}$ on $\varepsilon_0$ in $\ell_q$-norm. The related physical dependence measures are defined as
$$ \Omega_{t_0,q}^{(a)} := \sum_{t=t_0}^\infty \delta_{t,q}^{(a)},\quad \Omega_{t_0, q} := \max_{a\in [p]} \Omega_{t_0,q}^{(a)}. $$
Additionally, $\mathcal{F}_{t-t_0, t} := \sigma(\varepsilon_{t-t_0}, \ldots, \varepsilon_t)$ and $\mathcal{F}_t := \mathcal{F}_{-\infty,t}$. $t_0$-dependent approximating sequence of $X_t$ is defined as
$$ \tilde X_t^{(a)} := \EE[X_t^{(a)} \mid \mathcal{F}_{t-t_0, t}],~~ a \in [p],\quad \tilde X_t := \EE[X_t \mid \mathcal{F}_{t-t_0, t}],\quad t_0 \ge 0. $$

\begin{lemma}[Joint CLT of bilinear form]\label{lem:joint_clt}
Consider $n$ observations from a time series $\{Z_t\}_{t\in[n]}$ satisfying Assumption \ref{asn:summable}. Let $\alpha_{n, \ell} = \beta_{n, \ell} e^{-\i \ell \omega_j}$, where $j \in F_n$ and $\beta_{n, \ell} \in \RR$ with $\beta_{n, \ell} = \beta_{n, -\ell}$. For $u_k, v_k \in \RR^p$, $k \in [q]$, Define
$$ \Bcal_n(u_k, v_k) := \sum_{1 \le \ell, \ell' \le n} \alpha_{n,\ell - \ell'} (u_k^\top X_\ell) (v_k^\top X_{\ell'}), \text{ and } \sigma_n^2 := \kappa(\omega) \sum_{r, t = 1}^n \beta_{n, t - r}^2, $$
where $\kappa(x) = 2$ if $x/\pi \in \ZZ$ and 1 otherwise. Assume $\EE[X_t] = 0$, $\|X_0\|_4 < \infty$, $\Omega_{0, 4} < \infty$ and
\begin{align}
    & \max_{0\le t \le n} \beta_{n,t}^2 = o(\zeta_n^2), \text{ where }  \zeta_n^2 = \sum_{t = 1}^n \beta_{n,t}^2, \text{ and }
    n\zeta_n^2 = \bigO{\sigma_n^2},\\
    & \sum_{r = 1}^n \sum_{t = 1}^{n-1} \left|\sum_{\ell = 1 + r}^n \alpha_{n, r-\ell} \alpha_{n, t-\ell}\right|^2 = o(\sigma_n^4), \text{ and}\\
    & \sum_{t = 1}^n |\beta_{n,r} - \beta_{n,r-1}|^2 = o(\zeta_n^2).
\end{align}
Then for $\omega_j \in (0, \pi)$ and $c_k \in \RR$, $k \in [q]$,
$$ \frac{\sum_{k = 1}^q c_k \Bcal_n(u_k, v_k) - \EE[\sum_{k = 1}^q c_k \Bcal_n(u_k, v_k)]}{\sigma_n} \dconverge 2\pi(\zeta_1 + \i \zeta_2), $$
where $(\zeta_1, \zeta_2)^\top$ is a bivariate normal random variable with $\EE[\zeta_1] = 0$, $\EE[\zeta_2] = 0$, and
\begin{align*}
    \EE[\zeta_2^2] = & \frac{1}{2} \sum_{k_1, k_2 = 1}^q \left( c_{k_1} \overline{c}_{k_2} {f}^{[u_{k_1}, u_{k_2}]} {f}^{[v_{k_1}, v_{k_2}]} - \re(c_{k_1} c_{k_2} {f}^{[u_{k_1}, v_{k_2}]} {f}^{[u_{k_2}, v_{k_1}]})\right),\\
    \EE[\zeta_1^2] = & \frac{1}{2} \sum_{k_1, k_2 = 1}^q \left( c_{k_1} \overline{c}_{k_2} {f}^{[u_{k_1}, u_{k_2}]} {f}^{[v_{k_1}, v_{k_2}]} + \re(c_{k_1} c_{k_2} {f}^{[u_{k_1}, v_{k_2}]} {f}^{[u_{k_2}, v_{k_1}]})\right),\\
    \EE[\zeta_1 \zeta_2] = & \frac{1}{2} \sum_{k_1, k_2 = 1}^q \im(c_{k_1} c_{k_2} {f}^{[u_{k_1}, v_{k_2}]} {f}^{[u_{k_2}, v_{k_1}]}),
\end{align*}
where ${f}^{[u,v]} := u^\top f_j v$ for $u, v \in \RR^p$.
\end{lemma}

Lemma \ref{lem:joint_clt} is generalizes and gets inspirations from \citet[Lemma A.6]{wu2018asymptotic} that establishes CLT for the bilinear form $\sum_{1 \le \ell,\ell'\le n} \alpha_{n, \ell - \ell'} X_\ell^{(a)} X_{\ell'}^{(b)} $. Here we generalize that for deriving the joint asymptotic limit of finite collection of such bilinear forms.

For $u, v \in \CC^p$, definition of $\Bcal_n(\cdot, \cdot)$ can be extended as
\begin{align*}
    \Bcal_n(u, v) := & \Bcal_n(\re(u), \re(v)) - \i \Bcal_n(\im(u), \re(v)) \\
     + &\i \Bcal_n(\re(u), \im(v)) + \Bcal_n(\im(u), \im(v)),
\end{align*}
and similarly
$$f^{[u,v]} := f^{[\re(u), \re(v)]} - \i f^{[\im(u), \re(v)]} + \i f^{[\re(u), \im(v)]} + f^{[\im(u), \im(v)]}.$$

\begin{lemma}[CLT of bilinear form for complex vectors]\label{lem:cmplx_clt}
Suppose $u, v \in \CC^p$. Then under the conditions of Lemma \ref{lem:joint_clt}, 
\begin{equation}\label{eq:cmplx_clt}
    \frac{\Bcal_n(u,v) - \EE[\Bcal_n(u,v)]}{\sigma_n} \dconverge 2\pi(\zeta_1 + \i \zeta_2),
\end{equation}
where $(\zeta_1, \zeta_2)^\top$ is a bivariate normal random variable with $\EE[\zeta_1] = 0$, $\EE[\zeta_2] = 0$, and
\begin{align*}
    \EE[\zeta_1^2] = & \frac{1}{2}(f^{[u,u]} f^{[v,v]} + \re((f^{[u,v]})^2)),\\
    \EE[\zeta_2^2] = & \frac{1}{2}(f^{[u,u]} f^{[v,v]} - \re((f^{[u,v]})^2)),\\
    \EE[\zeta_1 \zeta_2] = & \frac{1}{2} \im((f^{[u,v]})^2),\\
\end{align*}
where ${f}^{[u,v]} := u^* f(\omega_j) v$ for $u, v \in \CC^p$.
\end{lemma}

The proof of Lemma \ref{lem:cmplx_clt} directly follows from Lemma \ref{lem:joint_clt} with $q = 4$ and $c_1 = 1$, $c_2 = -\i$, $c_3 = \i$, $c_4 = 1$.

\begin{lemma}[Variance expression for Dirichlet kernel] \label{lem:variance_dirichlet}
    In Lemma \ref{lem:joint_clt}, assume $\beta_{n,h} = D_m\left( \frac{2\pi h}{n} \right) \big/ (2\pi n (2m+1))$. Then $\sigma_n^2 = \sum_{t, r = 1}^n \beta_{n, t-r}^2 = \frac{1}{4\pi^2(2m+1)}$.
\end{lemma}

The proof of Lemma \ref{lem:joint_clt} follows arguments similar to those in \citet[Proof of Lemma A.6]{wu2018asymptotic}.

\subsection{Proof of Lemma \ref{lem:joint_clt}}\label{pf:joint_clt}
Assumption \ref{asn:summable} implies that $\Omega_{0,4} < \infty$. For $u \in \RR^q$, we define $X_t^{[u]} := u^\top X_t$ and $\tilde X_t^{[u]} = u^\top \tilde X_t$. Additionally, define 
$$A_{t,t_0}^{[u]} := \sum_{h = 0}^{t_0} \EE[\tilde X_{t+h}^{[u]} \mid \Fcal_t] e^{\i h\omega_j},\quad D_{t}^{[u]} := A_{t,t_0}^{[u]} - \EE[A_{t,t_0}^{[u]} \mid \Fcal_{t-1}].  $$
Following \citet{wu2018asymptotic}, $D_{t}^{[u]}$ is $\Fcal_{t-t_0, t}$-measurable and $\{D_t^{[u]} : t \in [n]\}$ is a $t_0$-dependent Martingale difference sequence. Define
$$ M_n^{[u, v]} := \sum_{t = 1}^n \overline{D}_t^{[u]} \sum_{\ell = 1}^{t-1} \alpha_{\ell - t} D_\ell^{[v]} $$ 
which is a $\Fcal_{t-t_0, t}$-measurable Martingale sequence. Following \citet[Lemmas A.1 and A.2]{wu2018asymptotic}, 
$$ \left\| \sum_{k=1}^q c_k \left( \sum_{1\le \ell < \ell' \le n} \alpha_{n, \ell - \ell'} X_{\ell}^{[u_k]} X_{\ell}^{[v_k]} - M_n^{[u_k, v_k]} \right) \right\| = o(\sigma_n^4). $$
Next, we show along \citet[Lemma A.6]{wu2018asymptotic} that using Bernstein's lemma, and showing that the conditions of Martingale CLT \citep[Theorem 3.2]{hall2014martingale} are met, the result can be proved. We denote the following quantities
$$ U_t^{[u]\dagger} := \sum_{\ell = \max\{t-4t_0+1, 1\}}^{t-1} \alpha_{n, \ell - t} D_{\ell}^{[u]},\quad U_t^{[u]\diamond} := \sum_{\ell = 1}^{t-4t_0} \alpha_{n, \ell - t} D_{\ell}^{[u]},$$
and the spectral density for the truncated process as $\tilde f^{[u,v]}(\omega) := \frac{1}{2\pi} \sum_{\ell = -t_0}^{t_0} e^{\i \ell \omega} \EE[\tilde X_0^{[u]} \tilde X_{\ell}^{[v]}]$. Since the linear combination is taken over finitely many $c_k$'s, the conditions
$$ \left\| \ \sum_{t=1}^n \overline{D}_t^{[u]} U_t^{[v]\dagger} \right\| \le C\sqrt{n} \max_{t\in [n]} |b_{n, t}|,\quad \text{and} \sum_{t = 1+4t_0}^n \|\overline{D}_t^{[u]} U_t^{v\diamond}\| = o(\sigma_n^4) $$
hold. Hence Lindeberg's condition is satisfied and the central limit theorem holds, such that
$$\sum_{k = 1}^q (M_n^{[u_k,v_k]} +  \overline{M}_n^{[u_k,v_k]}) / \sigma_n \xrightarrow{d} 2\pi(\tilde \zeta_1 + \i \tilde \zeta_2)$$ 
where $(\tilde \zeta_1, \tilde \zeta_2)^\top$ is bivariate mean-zero Gaussian random variable. It remains to show that the covariance matrix of  $(\tilde \zeta_1, \tilde \zeta_2)^\top$ is the same as that of $(\zeta_1, \zeta_2)^\top$, with ${f}^{[\cdot, \cdot]}$ replaced by $\tilde f^{[\cdot, \cdot]}$. Using $\EE[\cdot\mid \Fcal_{t-1}] = \sum_{s = 1}^{t_0} (\EE[\cdot \mid \Fcal_{t-s}] - \EE[\cdot \mid \Fcal_{t-s-1}]) + \EE[\cdot \mid \Fcal_{t-t_0}]$, we can show for $-t_0 \le s \le t_0-1$:
\begin{align*}
& \left\| \sum_{t = 1+4t_0}^n \Big( \EE\!\left[\left|\sum_{k=1}^{q} c_k \big( \overline{D}_t^{[u_k]} U_t^{[v_k]\diamond} + D_t^{[v_k]} \overline{U}_t^{[u_k]\diamond}\big)\right|^{2}\;\middle|\; \Fcal_{t-r} \right]\right. \\
&\qquad\qquad\left.- \EE\!\left[\left|\sum_{k=1}^{q} c_k \big( \overline{D}_t^{[u_k]} U_t^{[v_k]\diamond} + D_t^{[v_k]} \overline{U}_t^{[u_k]\diamond}\big)\right|^{2}\;\middle|\;\Fcal_{t-r-1}\right]\Big)\right\|\\
= & 4 k~ \max_{ k \in [q] }\ c_k \sum_{t = 1+4t_0}^n \|D_t^{[u_k]}\|_4^4 \|U_t^{v_k}\|_4^4 \le C n \zeta_n^2 = o(\sigma_n^4).
\end{align*}
And $D_t^{[u]}$ is $\Fcal_{t-t_0,t}$-measurable and $U_t^{[u]\diamond}$ is $\Fcal_{t-4t_0,t}$-measurable. Following \citet[Lemma A.6]{wu2018asymptotic}, $\EE[\overline{D}_t^{[u_{k_1}]} U_t^{[v_{k_1}]\diamond} D_t^{[u_{k_2}]} \overline{U}_t^{[v_{k_2}]\diamond} ] = U_t^{[v_{k_1}]\diamond} \overline{U}_t^{[v_{k_2}]\diamond} \EE[\overline{D}_t^{[u_{k_1}]} D_t^{[u_{k_2}]} ] $ and similarly for the other terms in the expression to follow:
\begin{align*}
& \EE\!\left[\left|\sum_{k=1}^{q} c_k \big( \overline{D}_t^{[u_k]} U_t^{[v_k]\diamond} + D_t^{[v_k]} \overline{U}_t^{[u_k]\diamond}\big)\right|^{2}\;\middle|\; \Fcal_{t-t_0} \right]\\
= & \sum_{t = 1+4t_0} \sum_{k_1, k_2 = 1}^q c_{k_1} \overline{c}_{k_2} \EE\left[ \left(\overline{D}_t^{[u_{k_1}]} U_t^{[v_{k_1}]\diamond} + D_t^{[v_{k_1}]} \overline{U}_t^{[u_{k_1}]\diamond}\right) \left(D_t^{[u_{k_2}]} \overline{U}_t^{[v_{k_2}]\diamond} + \overline{D}_t^{[v_{k_2}]} U_t^{[u_{k_2}]\diamond}\right) \;\middle|\; \Fcal_{t-t_0} \right]\\
= & \sum_{t=1+4t_0}^n \sum_{k_1,k_2=1}^q c_{k_1} \overline{c}_{k_2} \bigg[ U_t^{[v_{k_1}]\diamond} \overline{U}_t^{[v_{k_2}]\diamond} \EE[\overline{D}_t^{[u_{k_1}]} D_t^{[u_{k_2}]} ] + U_t^{[v_{k_1}]\diamond} U_t^{[u_{k_2}]\diamond} \EE[\overline{D}_t^{[u_{k_1}]} \overline{D}_t^{[v_{k_2}]} ] +  \\
& \hspace{100pt} \overline{U}_t^{[u_{k_1}]\diamond} \overline{U}_t^{[v_{k_2}]\diamond} \EE[D_t^{[v_{k_1}]} D_t^{[u_{k_2}]} ] + \overline{U}_t^{[u_{k_1}]\diamond} U_t^{[u_{k_2}]\diamond} \EE[D_t^{[v_{k_1}]} \overline{D}_t^{[v_{k_2}]} ] \bigg]
\end{align*}
We have $\EE[\overline{D}_t^{[u]} D_t^{[v]}] = 2\pi \tilde f^{[u,v]}(\omega)$, and $\|\sum_{t = 1+4t_0}^n U_t^{[u]\diamond} U_t^{[v]\diamond}\| = o_\PP(\sigma_n^2)$. Additionally,
$$ \EE[U_t^{[u]\diamond} \overline{U}_t^{[v]\diamond}] = \sum_{\ell = 1}^{t-4t_0} \beta_{n, \ell - t}^2 \EE[\overline{D}_t^{[u]} D_t^{[v]}]. $$
Thus the limiting variance is
\begin{align*}
& \frac{1}{\sigma_n^2}\sum_{t = 1+4t_0}^n \EE\!\left[\left|\sum_{k=1}^{q} c_k \big( \overline{D}_t^{[u_k]} U_t^{[v_k]\diamond} + D_t^{[v_k]} \overline{U}_t^{[u_k]\diamond}\big)\right|^{2}\;\middle|\; \Fcal_{t-1} \right] \\
\xrightarrow{\PP} &\ 4\pi^2 \sum_{k_1, k_2 = 1}^q c_{k_1} \overline{c}_{k_2} \tilde f^{[u_{k_1}, u_{k_2}]}(\omega_j) \tilde f^{[v_{k_1}, v_{k_2}]}(\omega_j).
\end{align*}
Similarly, the limiting pseudovariance is
\begin{align*}
& \frac{1}{\sigma_n^2}\sum_{t = 1+4t_0}^n \EE\!\left[\left(\sum_{k=1}^{q} c_k \big( \overline{D}_t^{[u_k]} U_t^{[v_k]\diamond} + D_t^{[v_k]} \overline{U}_t^{[u_k]\diamond}\big)\right)^{2}\;\middle|\; \Fcal_{t-1} \right] \\
\xrightarrow{\PP} &\ 4\pi^2 \sum_{k_1, k_2 = 1}^q c_{k_1} c_{k_2} \tilde f^{[u_{k_1}, v_{k_2}]}(\omega_j) \tilde f^{[u_{k_2}, v_{k_1}]}(\omega_j).
\end{align*}
and the proof is complete. \pfend

\subsection{Proof of Lemma \ref{lem:variance_dirichlet}}\label{pf:variance_dirichlet}

Since $D_m(x) = 1 + 2\sum_{h = 1}^m \cos(hx)$, we have
$$ D_m\left( \frac{2\pi (h+n)}{n} \right) = D_m\left( \frac{2\pi h}{n} + 2\pi \right) = D_m\left( \frac{2\pi h}{n} \right), $$
implying $\beta_{n,h+n} = \beta_{n,h}$ for every $h$. Thus we have exactly $n$ pairs of $t,r$ with $t - r \equiv h\mod n$. Therefore
\begin{align*}
    \sigma_n^2 
    = & \sum_{r,t=1}^n \beta_{n,t-r}^2 = n\sum_{h = 0}^{n-1} \beta_{n,h}^2 
    =  n \sum_{h=0}^{n-1} \left( \frac{D_m\left(\frac{2\pi h}{n}\right)}{2\pi n(2m+1)} \right)^2\\
    = & \frac{1}{4\pi^2 n(2m+1)^2} \sum_{h = 0}^{n-1} D_m\left( \frac{2\pi h}{n} \right)^2.
\end{align*}
Furthermore, 
\begin{align*}
    \sum_{h = 0}^{n-1} D_m\left( \frac{2\pi h}{n} \right)^2 = & \sum_{k, \ell = -m}^m \sum_{h = 0}^{n-1} e^{-\i (k+\ell) \frac{2\pi h}{n}}\\
    = & \sum_{k, \ell = -m}^m \# \{(k,\ell): -m \le k,\ell \le m, (k+\ell)/n \in \ZZ\} = n(2m+1).
\end{align*}
Therefore the final variance expression is $\sigma_n^2 = \frac{1}{4\pi^2 n(2m+1)^2} \times n(2m+1) = \frac{1}{4\pi^2(2m+1)}$. Hence the proof is complete. \pfend

\section{Proof of Theorem \ref{thm:consistency_variance}}\label{pf:consistency_variance}

\subsection{Consistency of plug-in estimator}

We denote the population analogues $\Phi_{j,k}^{(a,b)} := (\Theta_j)_{\cdot,a}^* f_k (\Theta_j)_{\cdot,b}$ and $\Xi_{j,k}^{(a,b)} := (\Theta_j)_{\cdot,a}^\top f_k (\Theta_j)_{\cdot,b}$. By \citet[Theorem 4.1 and Lemma B.4]{deb2024regularized} and under the assumptions of Theorem \ref{thm:consistency_variance},
$$ |\hat \Phi_{j,k}^{(a,b)} - \Phi_{j,k}^{(a,b)}| = \bigOP{\M^2 (\sqrt{d} \kt C_\gamma \thres + d \thres)}, $$
and $|\Phi_{j,k}^{(a,b)}| \le \frac{\M^2 \lf m}{2n} + \M = \bigO{1}$. Similar bounds hold for $|\hat \Xi_{j,k}^{(a,b)} - \Xi_{j,k}^{(a,b)}|$ and $\Xi_{j,k}^{(a,b)}$. By Lemma \ref{lem:consistency_oracle_hac} and \ref{lem:4th_cum},
\begin{align*}
    & \left|\frac{1}{(2m+1)^2} \sum_{k_1, k_2\in \wmj} \left(\Phi_{j,k}^{(a,a)} \Phi_{j,k_2}^{(b,b)} \indic\{k_1 = k_2\} + \Xi_{j,k_1}^{(a,b)} \overline{\Xi_{j,k_1}^{(a,b)}} \indic\left\{k_1 + k_2 \in n \ZZ\right\} \right)\right.\\
    & \hspace{10pt} -\var((\Theta_j)_{\cdot, a}^* W (\Theta_j)_{\cdot, b})\Big| = \bigO{\frac{d^2\M^4}{n(2m+1)}}.
\end{align*}
Therefore combining both bounds,
$$ |(\hat \sigma_j^2)^{(a,b)}-\var((\Theta_j)_{\cdot, a}^* W (\Theta_j)_{\cdot, b})| = \bigO{d\thres + \frac{d^2}{n(2m+1)}}. $$

\subsection{Consistency of HAC estimator}\label{sec:consistency_hac}

We denote $\hat S_n := (\hat \sigma_j^2)^{(a,b); \mathrm{HAC}} = \frac{1}{(2m+1)^2} \sum_{k_1, k_2 \in \wmj}^m W_{k_1, k_2} \hat y_{j,k_1}^{(a,b)} \overline{\hat y_{j,k_2}^{(a,b)}}$. We prove that $\hat S_n$ is consistent for its population analogue $S_n$, which in turn is consistent to $\var((\Theta_j)_{\cdot, a}^* W (\Theta_j)_{\cdot, b})$. Similarly, one can show that $(\hat{\delta}_j^2)^{(a,b);\mathrm{HAC}}$ is consistent for $\pvar\big((\Theta_j)_{\cdot,a}^* W, (\Theta_j)_{\cdot,b}\big)$; we omit the proof for brevity.

Next, we use \citet[Lemmas B.1 and B.3, and Theorem 4.1]{deb2024regularized}.
\begin{align*}
    \hat y_{j,k}^{(a,b)} - y_{j,k}^{(a,b)}
    = & \underbrace{((\hat \Theta_j)_{\cdot, a}-(\Theta_j)_{\cdot, a})^* (d_k d_k^{*}) ((\hat \Theta_j)_{\cdot, b}-(\Theta_j)_{\cdot, b})}_{(I)}\\
    & \hspace{10pt} - \underbrace{[(\hat \Theta_j)_{\cdot, a}^* \hat f_k (\hat \Theta_j)_{\cdot, b} - (\Theta_j)_{\cdot, a}^* f_k (\Theta_j)_{\cdot, b}]}_{(II)} + \bigO{\M^2 \left( \tnf + \frac{1}{4\pi n}\lf \right)}.
\end{align*}
Since under the conditions of Theorem \ref{thm:consistency_variance}, $\|(\hat \Theta_j)_{\cdot, a} - (\Theta_j)_{\cdot, a}\|_2 \le \sqrt{d}\|\hat \Theta_j - \Theta_j\|_\infty \le 2\kt C_\gamma \sqrt{d}\thres$,
$$ (I) = \bigOP{d \thres^2}, $$
and
$$ (II) = \bigOP{\M^2 \sqrt{d} \kt C_\gamma \thres + d\thres}. $$
Therefore with $\M, \kt, C_\gamma = \bigO{1}$,
$$ |\hat y_{j,k}^{(a,b)} - y_{j,k}^{(a,b)}| = \bigOP{d\thres}, $$
and
$$ |\hat S_n - S_n| = \bigOP{d\thres}. $$
By Lemma \ref{lem:consistency_oracle_hac} and Chebychev's inequality,
$$ |S_n - \var((\Theta_j)_{\cdot, a}^* W (\Theta_j)_{\cdot, b})| = \bigOP{\frac{1}{2m+1} + \frac{d^2}{n(2m+1)}}. $$
Combining both these bounds, we complete the proof. \pfend

\begin{lemma}[Consistency of oracle HAC estimator]\label{lem:consistency_oracle_hac}
    Denote 
    $$y_{j,k}^{(a,b)} := Y_{j,k}^{(a)} \overline{Y_{j,k}^{(b)}} - \EE[Y_{j,k}^{(a)} \overline{Y_{j,k}^{(b)}}]$$
    and let the oracle HAC estimator be denoted by
    $$S_n = \frac{1}{(2m+1)^2} \sum_{k_1, k_2 \in \wmj} W_{k_1, k_2} y_{j,k_1}^{(a,b)} \overline{y_{j,k_2}^{(a,b)}}$$ 
    where $W_{k_1, k_2} = \indic\{k_1 = k_2\} + \indic\{k_1 + k_2 \in n \ZZ\}$. Then under the conditions of Theorem \ref{thm:consistency_variance},
    $$ \EE[S_n] = \var((\Theta_j)_{\cdot, a}^* W (\Theta_j)_{\cdot, b}) + \bigO{\frac{1}{2m+1} + \frac{d^2}{n(2m+1)}},\quad \var(S_n) = \bigO{\frac{1}{(2m+1)^2} + \frac{d^4}{n}}. $$
\end{lemma}

\begin{lemma}[$r$\textsuperscript{th} order DFT cumulant; Theorem 4.3.2 \citep{brillinger2001time}]\label{lem:dft_cumulant}
    Under Assumption \ref{asn:8th_moment},
    $$ \cum\left(d_{k_1}^{(a_1)},\ldots, d_{k_r}^{(a_r)}\right) = (2\pi)^{\frac{r}{2}-1}n^{-\frac{r}{2}}\Delta_n\left(\sum_{j=1}^r \omega_j\right) f_{a_1,\ldots, a_r}(\omega_1,\ldots, \omega_{r-1}) + \bigO{n^{-\frac{r}{2}}}, $$
    where $f_{a_1,\ldots a_{r}}$ denotes the $r$\textsuperscript{th} order spectral density of $X_t$ defined as
    $$ f_{a_1, \ldots, a_r}(\omega_1,\ldots, \omega_{r-1}) := \frac{1}{(2\pi)^{r-1}} \sum_{h_1, \ldots, h_{r-1} \in \ZZ} \cum(X_0^{(a_1)}, X_{h_1}^{(a_2)}, \ldots, X_{h_{r-1}}^{(a_r)}) e^{-\i(h_1 \omega_1 + \ldots + h_{r-1} \omega_{r-1})}, $$
    and
    $$\Delta_n(x) = \sum_{t = 0}^{n-1} e^{-\i xt} =
    \begin{cases}
    n & \text{ if } x \equiv 0\ (\mathrm{mod}\ 2\pi), \text{ and}\\
    0 & \text{ if } x = \frac{2\pi s}{n} \text{ for } s \not \equiv 0\ (\mathrm{mod}\ 2\pi).
    \end{cases}
    $$
\end{lemma}

\begin{lemma}[Bound on 4\textsuperscript{th} cumulant of the scaled DFT]\label{lem:4th_cum}
    Under Assumptions \ref{asn:summable} and \ref{asn:8th_moment}, and $m \precsim n/(\lf \M)$,
    \begin{align*}
        \cov\left(y_{j,k_1}^{(a,b)}, y_{j,k_2}^{(a,b)}\right) = & ~\cum\left(y_{j,k_1}^{(a,b)}, \overline{y_{j,k_2}^{(a,b)}}\right) = (\Theta_j)_{\cdot, a}^* f_{k_1} (\Theta_j)_{\cdot, a} \times (\Theta_j)_{\cdot, b}^* f_{k_1} (\Theta_j)_{\cdot, b} \indic\{k_1 = k_2\}\\
        + & (\Theta_j)_{\cdot, a}^* f_{k_1} \overline{(\Theta_j)_{\cdot, b}} \times (\Theta_j)_{\cdot, a}^\top f_{k_1} (\Theta_j)_{\cdot, b} \indic\{k_1 + k_2 \in n\ZZ\} + \bigO{\frac{d^2\M^4}{n}}\\
        = &~ \bigO{W_{k_1, k_2}\M^2 +\frac{d^2\M^4}{n}}.
    \end{align*}
\end{lemma}

\subsubsection{Proof of Lemma \ref{lem:consistency_oracle_hac}: Consistency of oracle HAC estimator}
We can expand the population variance as
\begin{align*}
    \var((\Theta_j)_{\cdot, a}^* W (\Theta_j)_{\cdot, b}) = & \frac{1}{(2m+1)^2} \cov\left( \sum_{k \in \wmj} Y_{j,k}^{(a)} \overline{Y_{j,k}^{(b)}}, \sum_{k \in \wmj} Y_{j,k}^{(a)} \overline{Y_{j,k}^{(b)}} \right)\\
    = & \frac{1}{(2m+1)^2} \sum_{k_1, k_2 \in \wmj}  \cov\left( y_{j,k_1}^{(a,b)},  y_{j,k_2}^{(a,b)} \right).
\end{align*}
Taking expectation of the variance estimator,
$$ \EE[S_n] = \frac{1}{(2m+1)^2} \sum_{k_1, k_2 \in \wmj} W_{k_1, k_2} \cov\left( y_{j,k_1}^{(a,b)},  y_{j,k_2}^{(a,b)} \right). $$
Hence using Lemma \ref{lem:4th_cum},
\begin{align*}
    \EE[S_n] - \var((\Theta_j)_{\cdot, a}^* W (\Theta_j)_{\cdot, b}) = & \frac{1}{(2m+1)^2}\sum_{k \in \wmj} \cov\left( y_{j,k}^{(a,b)},  y_{j,-k}^{(a,b)} \right)\\
    = & \bigO{\frac{1}{2m+1} + \frac{d^2}{n(2m+1)}}.
\end{align*}
Next we expand the variance term as
\begin{align*}
    \var(S_n) = & \frac{1}{(2m+1)^4} \sum_{k_1, k_2, k_3, k_4 \in \wmj} \cov\left( y_{j,k_1}^{(a,b)} \overline{y_{j,k_2}^{(a,b)}}, y_{j,k_3}^{(a,b)} \overline{y_{j,k_4}^{(a,b)}} \right)\\
    = & \frac{1}{(2m+1)^4} \sum_{k_1, k_2, k_3, k_4 \in \wmj}\left[ \cum\left(y_{j,k_1}^{(a,b)}, \overline{y_{j,k_3}^{(a,b)}}\right) \cum\left(\overline{y_{j,k_2}^{(a,b)}}, y_{j,k_4}^{(a,b)}\right)  \right.\\
    & \left. + \cum\left(y_{j,k_1}^{(a,b)}, y_{j,k_4}^{(a,b)}\right) \cum\left(\overline{y_{j,k_2}^{(a,b)}}, \overline{y_{j,k_3}^{(a,b)}}\right) + \cum\left(y_{j,k_1}^{(a,b)}, \overline{y_{j,k_2}^{(a,b)}}, \overline{y_{j,k_3}^{(a,b)}}, y_{j,k_4}^{(a,b)}\right) \right]
\end{align*}
Using Lemma \ref{lem:4th_cum}, the first term in the sum is 
$$ \cum\left(y_{j,k_1}^{(a,b)}, \overline{y_{j,k_3}^{(a,b)}}\right) \cum\left(\overline{y_{j,k_2}^{(a,b)}}, y_{j,k_4}^{(a,b)}\right) = W_{k_1, k_3} W_{k_2, k_4} \bigO{1} + \bigO{\frac{d^2}{n} + \frac{d^4}{n^2}}. $$
Hence the sum of the first term contains at most $4(2m+1)^2$-many non-zero $\bigO{1}$ terms. The second sum can be bounded similarly. For bounding the last term we use Lemma \ref{lem:leonov}. The leading terms in the sum come from connected pairings of the linear DFT factors that can be simplified as
\begin{align*}
    & \cov((\Theta_j)_{\cdot, a}^* d_{k_1}, (\Theta_j)_{\cdot, b}^* d_{k_2}) \times  \cov((\Theta_j)_{\cdot, a}^* d_{k_2}, (\Theta_j)_{\cdot, b}^* d_{k_3}) \\
    & \hspace{15pt} \times \cov((\Theta_j)_{\cdot, a}^* d_{k_3}, (\Theta_j)_{\cdot, b}^* d_{k_4}) \times \cov((\Theta_j)_{\cdot, a}^* d_{k_4}, (\Theta_j)_{\cdot, b}^* d_{k_1})\\
    = & ((\Theta_j)_{\cdot, a}^* f_{k_1} (\Theta_j)_{\cdot, b})^4 \indic\{k_1 = k_2 = k_3 = k_4\}.
\end{align*}
Thus the sum of the last term contributes to at most $(2m+1)$-many non zero $\bigO{1}$ terms. Additionally using Lemma \ref{lem:leonov} and \ref{lem:dft_cumulant}, higher order cumulants contribute in the sum with with non-zero terms $\bigO{n^{1-\frac{r}{2}}}$ with $r \ge 4$. Therefore using Lemma \ref{lem:leonov} and \ref{lem:sparse_bound},
$$ \frac{1}{(2m+1)^4}\sum_{k_1, k_2, k_3, k_4 \in \wmj} \cum\left(y_{j,k_1}^{(a,b)}, \overline{y_{j,k_2}^{(a,b)}}, \overline{y_{j,k_3}^{(a,b)}}, y_{j,k_4}^{(a,b)}\right) = \bigO{\frac{1}{(2m+1)^3} + \frac{d^4}{n}}. $$
Combining the sums of all cumulants we obtain
$$ \var(S_n) = \bigO{\frac{1}{(2m+1)^2} + \frac{d^4}{n}}. $$
Hence the proof is complete. \pfend

\subsubsection{Proof of Lemma \ref{lem:4th_cum}: Bound on 4\textsuperscript{th} cumulant of the scaled DFT}\label{pf:4th_cum}

Lemma \ref{lem:4th_cum} implies,
\begin{align*}
    \cov\left(y_{j,k_1}^{(a,b)}, y_{j,k_2}^{(a,b)}\right) & = \cum\left( Y_{j,k_1}^{(a)} \overline{Y_{j,k_1}^{(b)}}, \overline{Y_{j, k_2}^{(a)}} Y_{j, k_2}^{(b)} \right)\\
    = & \cum\left(Y_{j,k_1}^{(a)}, \overline{Y_{j,k_2}^{(a)}}\right) \cum\left(\overline{Y_{j,k_1}^{(b)}}, Y_{j,k_2}^{(b)} \right) \\
     + &\cum\left(Y_{j,k_1}^{(a)}, Y_{j,k_2}^{(b)}\right) \cum\left(\overline{Y_{j,k_1}^{(b)}}, \overline{Y_{j,k_2}^{(a)}} \right)  + \cum\left(Y_{j,k_2}^{(a)}, Y_{j,k_2}^{(a)}, \overline{Y_{j,k_1}^{(b)}}, \overline{Y_{j,k_2}^{(b)}} \right)
\end{align*}
From Lemma \ref{lem:dft_cumulant} with $r = 2$ and under Assumption \ref{asn:summable},
$$ \cov\left(Y_{j,k_1}^{(a)}, Y_{j,k_2}^{(b)}\right) = \cum\left(Y_{j,k_1}^{(a)}, \overline{Y_{j,k_2}^{(b)}}\right) = (\Theta_j)_{\cdot, a}^* f_{k_1} (\Theta_j)_{\cdot, b} \indic\{k_1 = k_2\} + \bigO{\frac{\M^2}{n}}. $$
Similarly,
$$ \cum\left(Y_{j,k_1}^{(a)}, Y_{j,k_2}^{(b)}\right) = (\Theta_j)_{\cdot, a}^* f_{k_1} \overline{(\Theta_j)_{\cdot, b}} \indic\{k_1 + k_2 \in n\ZZ\} + \bigO{\frac{\M^2}{n}}. $$
Additionally for $m \precsim n/(\lf \M)$ and $k \in \wmj$,
$$ |(\Theta_j)_{\cdot, a}^* f_k (\Theta_j)_{\cdot, b}| \le |(\Theta_j)_{\cdot, a}^*(f_k - f_j)(\Theta_j)_{\cdot, b}| + |(\Theta_j)_{a, b}| \le \frac{\M^2 \lf m}{2n} + \M = \bigO{\M}. $$
Again using Lemma \ref{lem:dft_cumulant} and under Assumption \ref{asn:8th_moment}, the 4\textsuperscript{th} cumulant expression is similarly
$$ \cum\left(Y_{j,k_1}^{(a)}, Y_{j,k_2}^{(a)}, \overline{Y_{j,k_1}^{(b)}}, \overline{Y_{j,k_2}^{(b)}} \right) = \bigO{d^{2} n^{-1} \M^4}. $$
Thus combining the three cumulant bounds, we complete the proof. \pfend

\begin{lemma}[Leonov--Shiryaev cumulant formula
{\citep{leonov1959method}; {\citep[Theorem 2.3.2]{brillinger2001time}}}]
\label{lem:leonov}
Let $\{X_{i,j}: 1\le i\le r,\ 1\le j\le q_i\}$ be random variables
with all required moments finite, and define
\[
Y_i=\prod_{j=1}^{q_i} X_{i,j},\qquad i=1,\ldots,r.
\]
Then
\[
\cum(Y_1,\ldots,Y_r)
=
\sum_{\pi\in\mathcal P_{\mathrm{conn}}}
\prod_{B\in\pi}
\cum\left( X_{i,j}:(i,j)\in B\right),
\]
where $\mathcal P_{\mathrm{conn}}$ denotes the collection of all partitions
$\pi$ of the index set $\{(i,j):1\le i\le r,\ 1\le j\le q_i\}$ that are connected with respect to the groups $G_i=\{(i,j):1\le j\le q_i\},\qquad i=1,\ldots,r$.
That is, $\pi\in\mathcal P_{\mathrm{conn}}$ if the graph on vertices $\{1,\ldots,r\}$, with an edge between $i$ and $i'$ whenever some block $B\in\pi$ intersects both $G_i$ and $G_{i'}$, is connected.
\end{lemma}

\begin{lemma}[Sparse contraction bound]
\label{lem:sparse_bound}
Let $Z_i = v_i^* d_{k_i}$ or $Z_i = d_{k_i}^* v_i$, $i=1,\dots,r$,
where $\|v_i\|_2 \le M$ and each $v_i$ is $d$-sparse.
Assume that for each $r\le 8$,
\[
\left|
\cum(d_{k_1}^{(a_1)},\dots,d_{k_r}^{(a_r)})
\right|
\le K_r n^{1-r/2}
\]
uniformly over indices and frequencies.
Then
\[
\left|
\cum(Z_1,\dots,Z_r)
\right|
\le C\, d^{r/2} M^r n^{1-r/2},
\]
where $C$ depends only on $K_r$.
\end{lemma}

\begin{remark}[Dimension factor in the sparse contraction bound]
The factor $d^{r/2}$ in Lemma \ref{lem:sparse_bound} arises from the bound $\|v_i\|_1 \le \sqrt{d}\|v_i\|_2 \le \sqrt{d}M$.
\end{remark}

\begin{lemma}[Cumulant representation of covariance {\citep[Chapter 3.2]{peccati2011wiener}}]\label{lem:cumulant} 
Suppose that the random variables $X_1, X_2, X_3, X_4\in\mathbb{C}$ have zero mean with $\EE[|X_1 X_2 X_3 X_4|] < \infty$, then
\begin{align*}
    & \cov(X_1 X_2, X_3 X_4) = \cum(X_1 X_2, \overline{X_3} \overline{X_4})\\
    =~ & \cum(X_1, \overline{X_3}) \cum(X_2,\overline{X_4}) + \cum(X_1, \overline{X_4}) \cum(X_2,\overline{X_3}) + \cum(X_1, X_2, \overline{X_3}, \overline{X_4}).
\end{align*}
\end{lemma}

\section{Additional simulation results}\label{app:additional_sim}

\subsection{Additional figures}

\begin{figure}[H]
\centering
\includegraphics[width=0.4\textwidth]{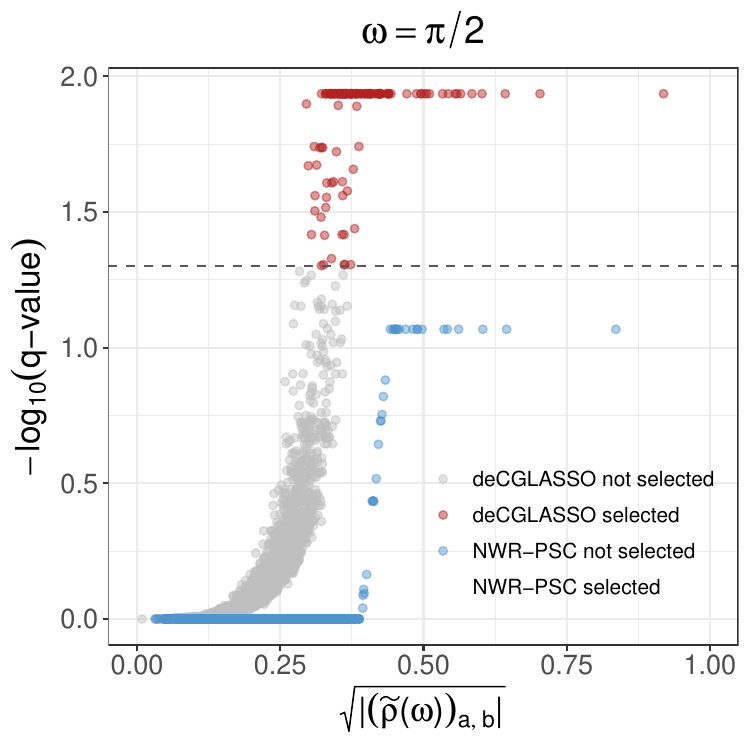}\hspace{10pt}
\includegraphics[width=0.4\textwidth]{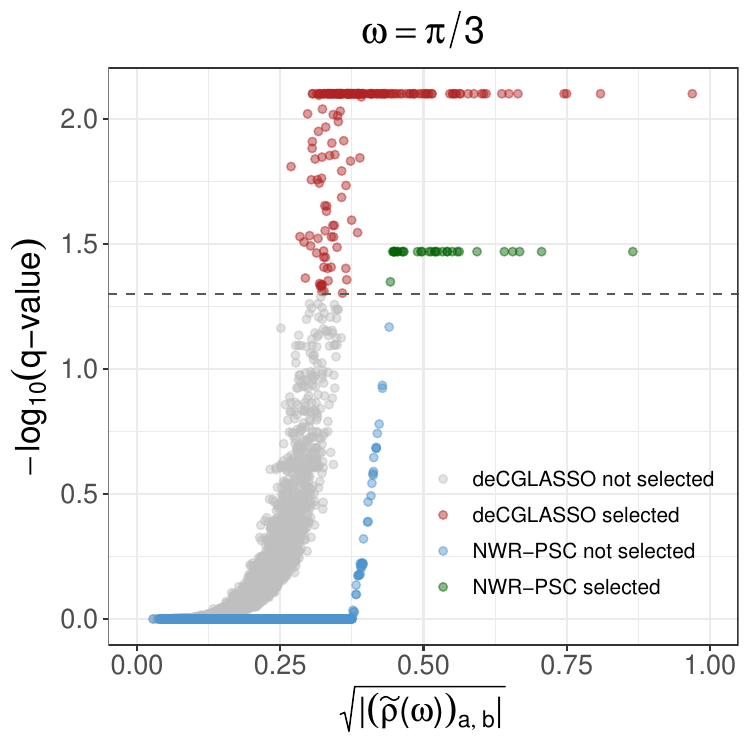}\\
\vspace{20pt}

\includegraphics[width=0.4\textwidth]{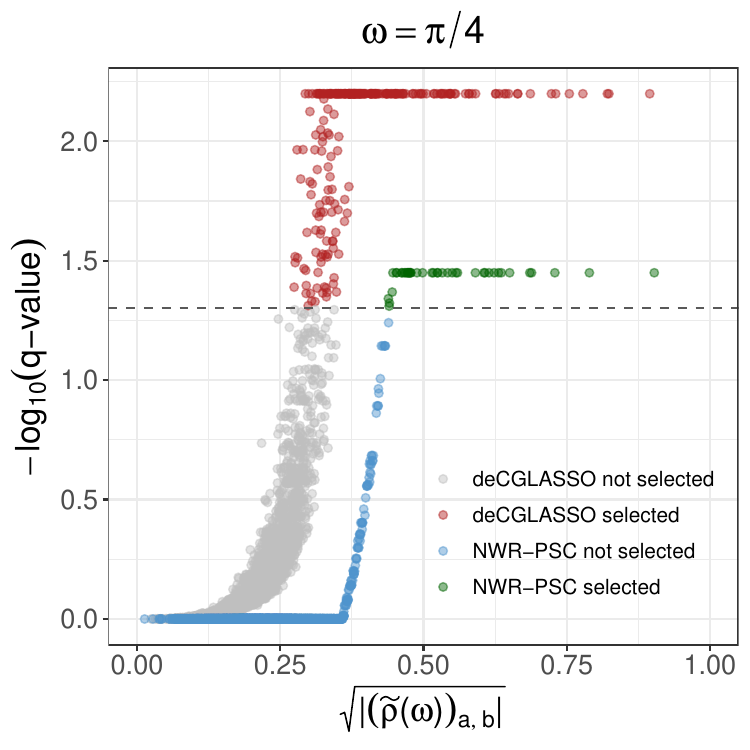}
\includegraphics[width=0.4\textwidth]{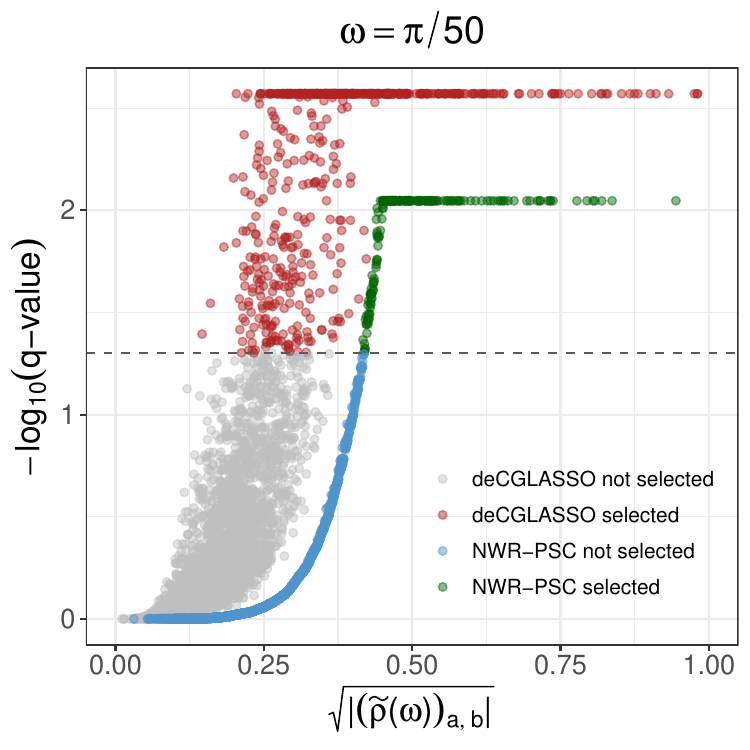}
\caption{\textbf{Additional volcano plots of deCGLASSO and NWR-PSC for a second representative subject at $\omega=\pi/2,\pi/3,\pi/4,$ and $\pi/50$.} Each point is a candidate edge $(a,b)$, with $x$-axis $\sqrt{|(\widetilde\rho(\omega))_{a,b}|}$ and $y$-axis $-\log_{10}(q^{\mathrm{thr}}_{a,b})$. The dashed line marks $-\log_{10}(0.05)\approx 1.301$, corresponding to the nominal FDR level $\alpha=0.05$.
}
\label{fig:volcano2}
\end{figure}

\begin{figure}[H]
\centering
\includegraphics[width=0.4\textwidth, trim={0.5in 0.1in 0in 0.25in}]{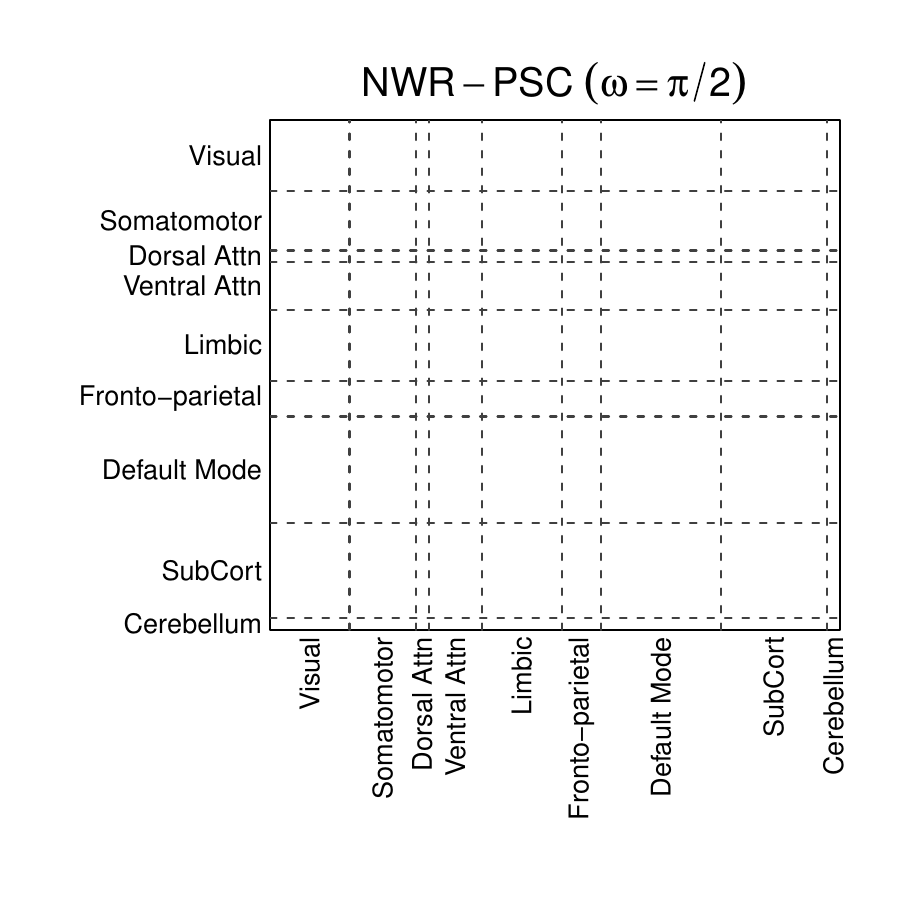}
\includegraphics[width=0.4\textwidth, trim={0.45in 0.3in 0.25in 0.25in}]{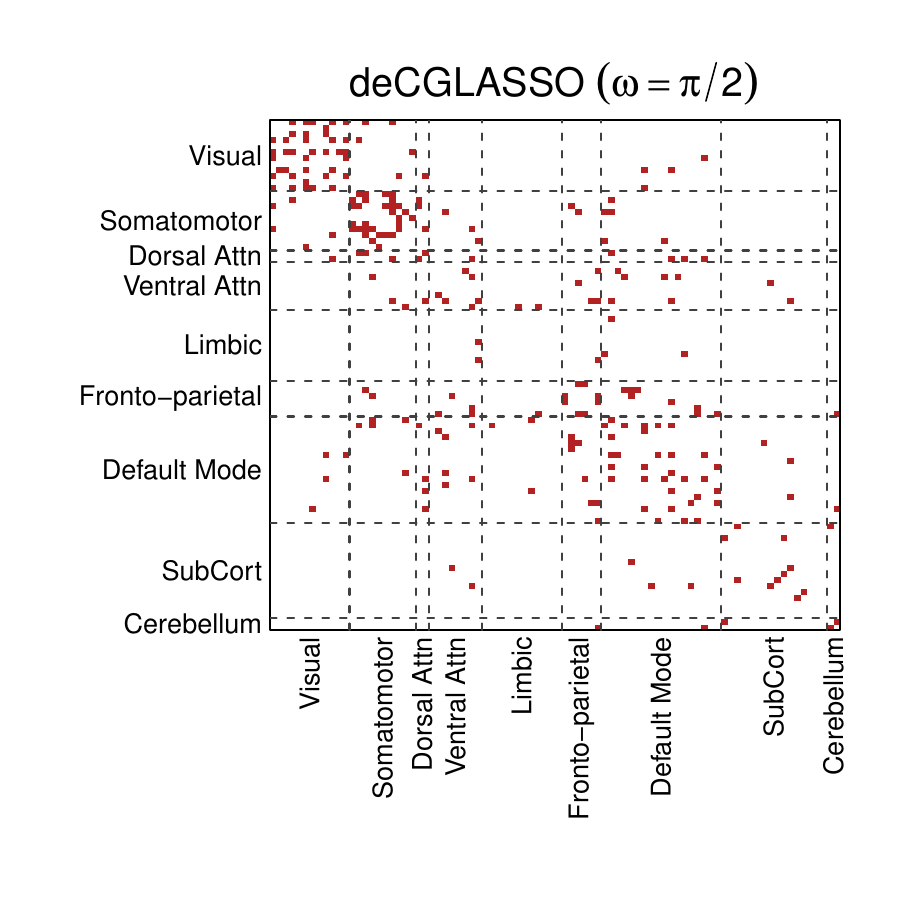}\\

\includegraphics[width=0.4\textwidth, trim={0.5in 0.3in 0.25in 0.25in}]{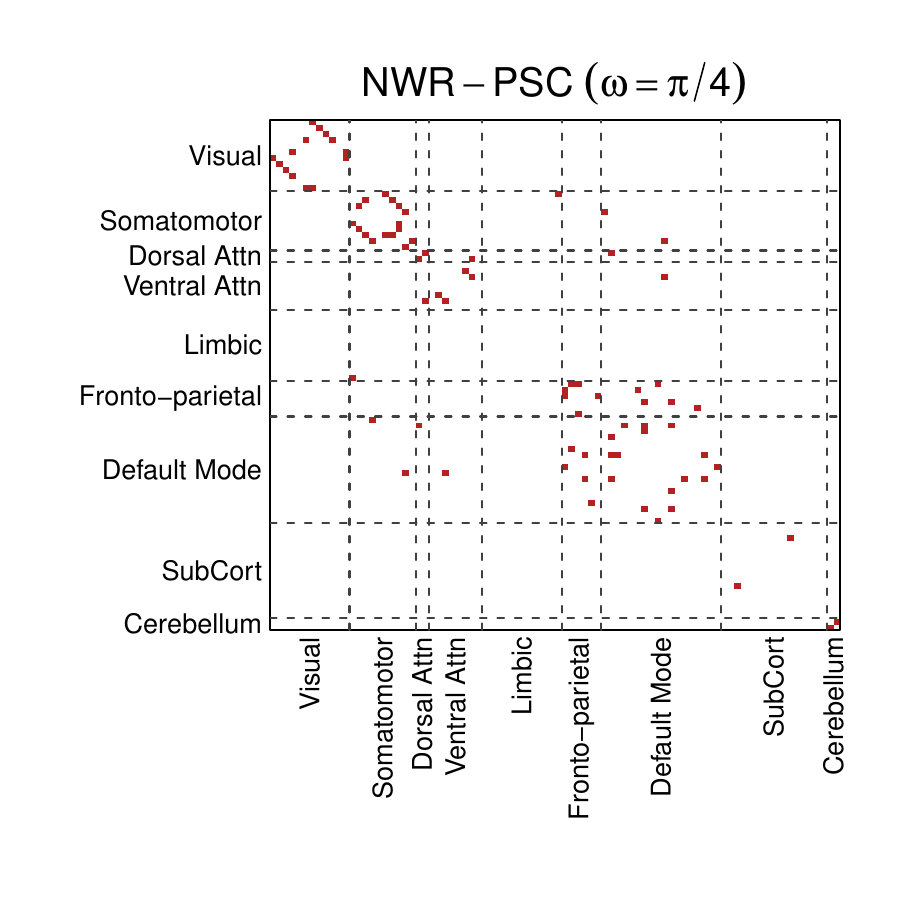}
\includegraphics[width=0.4\textwidth, trim={0.5in 0.3in 0.25in 0.25in}]{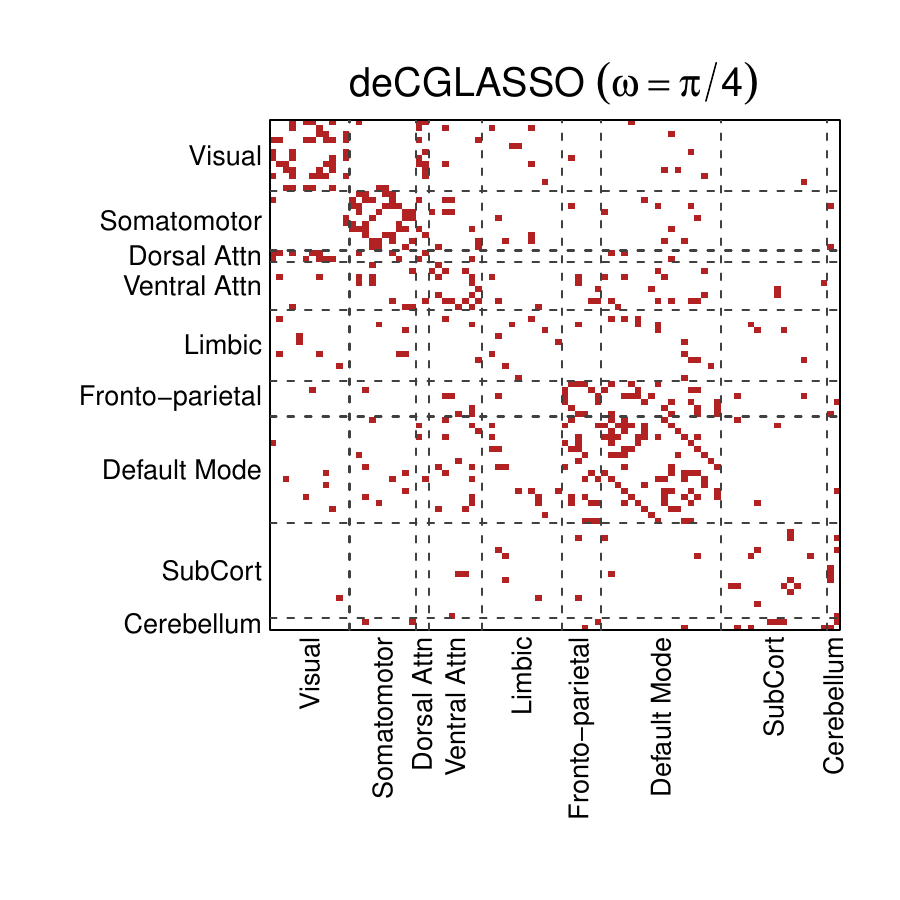}\\

\includegraphics[width=0.4\textwidth, trim={0.5in 0.3in 0.25in 0.25in}]{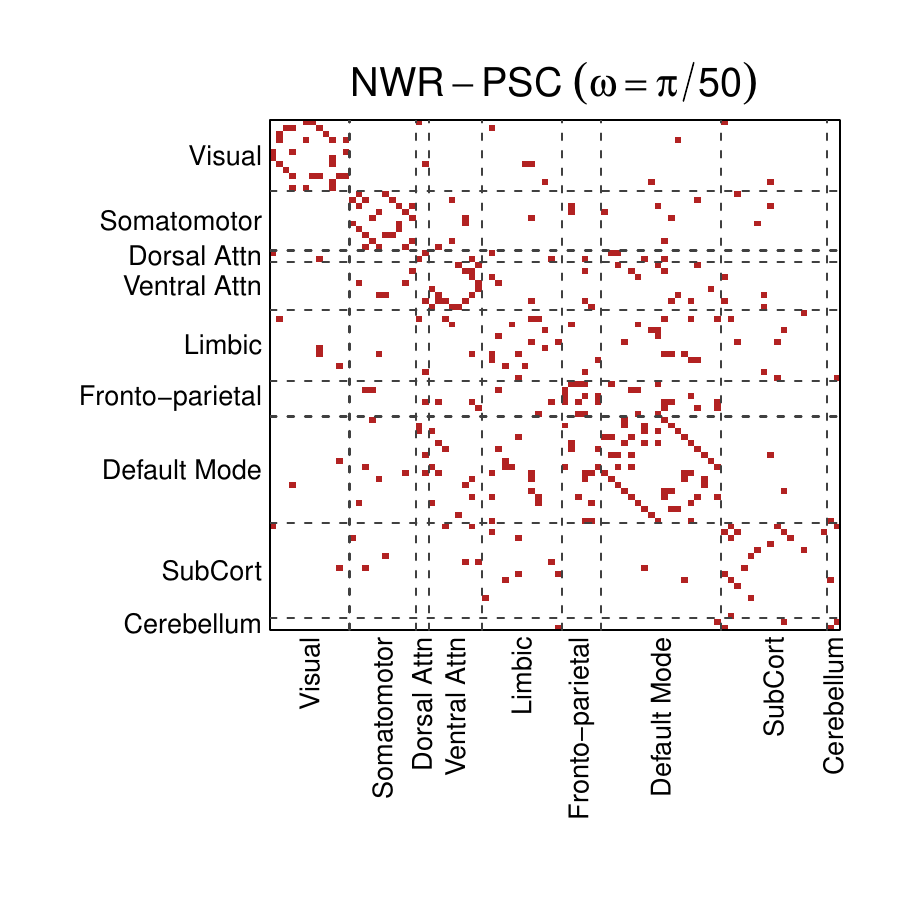}
\includegraphics[width=0.4\textwidth, trim={0.5in 0.3in 0.25in 0.25in}]{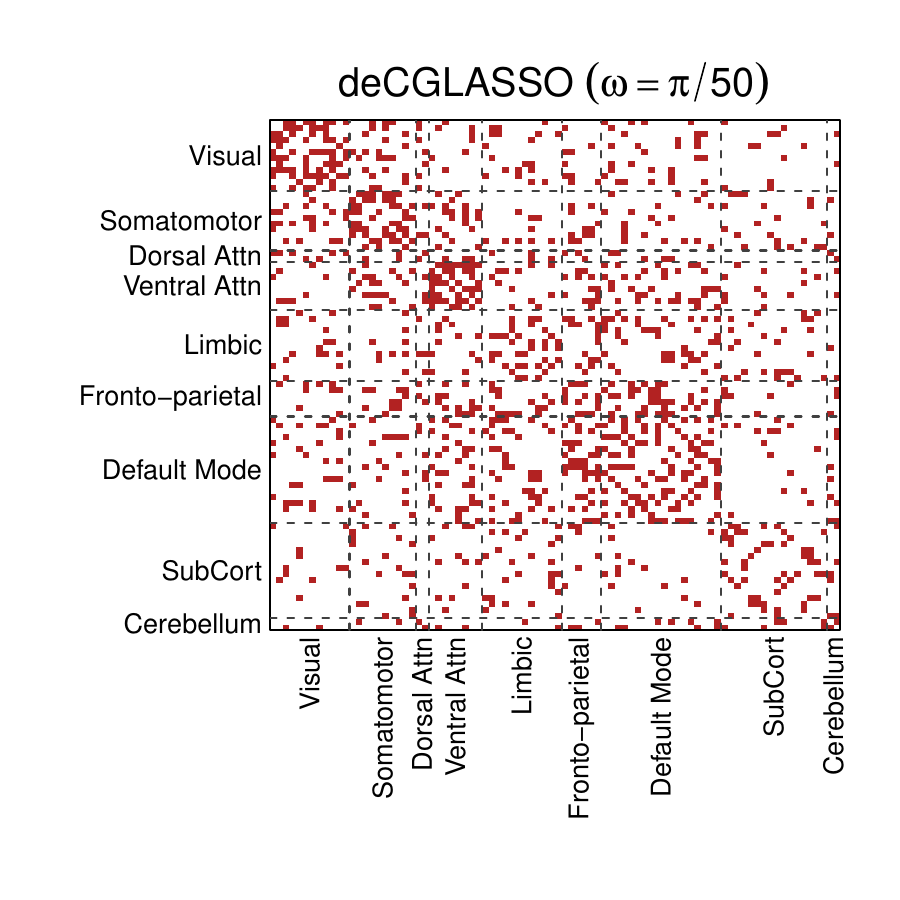}
\caption{\textbf{Additional heatmaps of significant edge adjacency matrices for a second representative subject using NWR-PSC and deCGLASSO at $\omega=\pi/2,\pi/4,$ and $\pi/50$.} Red entries indicate the significant entries selected by the FDR thresholding procedure; white entries indicate non-significant entries.}
\label{fig:heatmap2}
\end{figure}

\subsection{Simulation results on different data generating models}\label{app:more_table}

\begin{table}[H]
\centering
\small
\resizebox{0.95\columnwidth}{!}{%
\begin{tabular}{ccccccc}
\toprule
$p$ & $n$ & Method 
& $\textrm{Avgcov}_{j}(E)$ 
& $\textrm{Avgarea}_{j}(E)$ 
& $\textrm{Avgcov}_{j}(E^c)$ 
& $\textrm{Avgarea}_{j}(E^c)$ \\
\midrule
\multirow{12}{*}{50} 
& \multirow{3}{*}{400} 
& Pop     & 0.980 (0.015) & 21.705 (0.000) & 0.977 (0.006) & 16.227 (0.000) \\
& & Plug-in & 0.924 (0.026) & 18.734 (1.271) & 0.964 (0.007) & 13.604 (0.885) \\
& & HAC     & 0.900 (0.030) & 15.843 (0.944) & 0.934 (0.008) & 11.558 (0.537) \\
\cline{2-3}
& \multirow{3}{*}{900}
& Pop     & 0.978 (0.015) & 14.588 (0.000) & 0.977 (0.005) & 10.907 (0.000) \\
& & Plug-in & 0.917 (0.030) & 11.902 (0.597) & 0.957 (0.007) & 8.756 (0.431) \\
& & HAC     & 0.902 (0.028) & 10.849 (0.470) & 0.938 (0.008) & 7.994 (0.304) \\
\cline{2-3}
& \multirow{3}{*}{1600}
& Pop     & 0.971 (0.016) & 10.986 (0.000) & 0.977 (0.006) & 8.214 (0.000) \\
& & Plug-in & 0.907 (0.026) & 8.805 (0.405) & 0.953 (0.007) & 6.501 (0.294) \\
& & HAC     & 0.897 (0.029) & 8.243 (0.335) & 0.940 (0.007) & 6.092 (0.209) \\
\cline{2-3}
& \multirow{3}{*}{2500}
& Pop     & 0.961 (0.018) & 8.811 (0.000) & 0.978 (0.005) & 6.587 (0.000) \\
& & Plug-in & 0.896 (0.029) & 6.922 (0.289) & 0.951 (0.007) & 5.129 (0.210) \\
& & HAC     & 0.885 (0.031) & 6.610 (0.240) & 0.941 (0.007) & 4.890 (0.151) \\
\cline{1-3}
\multirow{12}{*}{150} 
& \multirow{3}{*}{400} 
& Pop     & 0.981 (0.008) & 21.685 (0.000) & 0.978 (0.003) & 16.257 (0.000) \\
& & Plug-in & 0.928 (0.015) & 18.680 (0.774) & 0.965 (0.003) & 13.590 (0.546) \\
& & HAC     & 0.901 (0.015) & 15.883 (0.510) & 0.935 (0.002) & 11.567 (0.342) \\
\cline{2-3}
& \multirow{3}{*}{900}
& Pop     & 0.977 (0.009) & 14.575 (0.000) & 0.979 (0.002) & 10.927 (0.000) \\
& & Plug-in & 0.918 (0.015) & 11.799 (0.362) & 0.958 (0.002) & 8.690 (0.259) \\
& & HAC     & 0.902 (0.016) & 10.739 (0.272) & 0.940 (0.003) & 7.946 (0.177) \\
\cline{2-3}
& \multirow{3}{*}{1600}
& Pop     & 0.970 (0.009) & 10.976 (0.000) & 0.979 (0.002) & 8.229 (0.000) \\
& & Plug-in & 0.905 (0.016) & 8.727 (0.251) & 0.955 (0.003) & 6.459 (0.183) \\
& & HAC     & 0.894 (0.017) & 8.187 (0.207) & 0.942 (0.002) & 6.063 (0.132) \\
\cline{2-3}
& \multirow{3}{*}{2500}
& Pop     & 0.961 (0.010) & 8.803 (0.000) & 0.979 (0.002) & 6.599 (0.000) \\
& & Plug-in & 0.896 (0.016) & 6.909 (0.163) & 0.953 (0.003) & 5.129 (0.119) \\
& & HAC     & 0.887 (0.016) & 6.572 (0.142) & 0.943 (0.002) & 4.888 (0.088) \\
\bottomrule
\end{tabular}%
}
\caption{\textbf{Additional average coverage probabilities and coverage areas over $E$ and $E^c$ for the VAR(1) model at $\omega=\pi/2$.} The reported values are averages over 200 repetitions (standard deviations in parentheses).}
\label{tab:sim_var1_nonzero_appendix}
\end{table}

\begin{table}[H]
\centering
\small
\resizebox{0.8\columnwidth}{!}{%
\begin{tabular}{ccccccc}
\toprule
$p$ & $n$ & Method 
& $\textrm{Avgcov}_{j}(E)$ 
& $\textrm{Avgarea}_{j}(E)$ 
& $\textrm{Avgcov}_{j}(E^c)$ 
& $\textrm{Avgarea}_{j}(E^c)$ \\
\midrule
\multirow{12}{*}{50} 
& \multirow{3}{*}{400} 
& Pop     & 0.923 (0.026) & 1.394 (0.000) & 0.998 (0.002) & 1.048 (0.000) \\
& & Plug-in & 0.436 (0.067) & 0.454 (0.066) & 0.947 (0.010) & 0.147 (0.020) \\
& & HAC     & 0.361 (0.069) & 0.425 (0.069) & 0.930 (0.008) & 0.134 (0.021) \\
\cline{2-3}
& \multirow{3}{*}{900}
& Pop     & 0.902 (0.034) & 0.937 (0.000) & 0.999 (0.001) & 0.705 (0.000) \\
& & Plug-in & 0.369 (0.062) & 0.251 (0.027) & 0.949 (0.008) & 0.105 (0.010) \\
& & HAC     & 0.325 (0.058) & 0.232 (0.029) & 0.931 (0.008) & 0.097 (0.010) \\
\cline{2-3}
& \multirow{3}{*}{1600}
& Pop     & 0.905 (0.034) & 0.706 (0.000) & 1.000 (0.001) & 0.531 (0.000) \\
& & Plug-in & 0.364 (0.067) & 0.178 (0.011) & 0.951 (0.007) & 0.087 (0.005) \\
& & HAC     & 0.330 (0.065) & 0.165 (0.011) & 0.934 (0.008) & 0.081 (0.005) \\
\cline{2-3}
& \multirow{3}{*}{2500}
& Pop     & 0.906 (0.034) & 0.566 (0.000) & 1.000 (0.000) & 0.426 (0.000) \\
& & Plug-in & 0.385 (0.070) & 0.144 (0.005) & 0.951 (0.007) & 0.078 (0.003) \\
& & HAC     & 0.360 (0.067) & 0.136 (0.006) & 0.936 (0.007) & 0.074 (0.003) \\
\cline{1-3}
\multirow{12}{*}{100} 
& \multirow{3}{*}{400} 
& Pop     & 0.911 (0.020) & 1.397 (0.000) & 0.999 (0.001) & 1.054 (0.000) \\
& & Plug-in & 0.391 (0.051) & 0.438 (0.044) & 0.950 (0.006) & 0.141 (0.013) \\
& & HAC     & 0.324 (0.049) & 0.412 (0.046) & 0.933 (0.004) & 0.129 (0.013) \\
\cline{2-3}
& \multirow{3}{*}{900}
& Pop     & 0.881 (0.027) & 0.939 (0.000) & 1.000 (0.000) & 0.709 (0.000) \\
& & Plug-in & 0.310 (0.042) & 0.240 (0.020) & 0.952 (0.005) & 0.100 (0.008) \\
& & HAC     & 0.268 (0.038) & 0.220 (0.020) & 0.934 (0.004) & 0.092 (0.007) \\
\cline{2-3}
& \multirow{3}{*}{1600}
& Pop     & 0.876 (0.028) & 0.707 (0.000) & 1.000 (0.000) & 0.534 (0.000) \\
& & Plug-in & 0.284 (0.041) & 0.167 (0.009) & 0.953 (0.003) & 0.081 (0.004) \\
& & HAC     & 0.256 (0.041) & 0.155 (0.009) & 0.937 (0.003) & 0.076 (0.004) \\
\cline{2-3}
& \multirow{3}{*}{2500}
& Pop     & 0.887 (0.029) & 0.567 (0.000) & 1.000 (0.000) & 0.428 (0.000) \\
& & Plug-in & 0.308 (0.051) & 0.133 (0.004) & 0.953 (0.003) & 0.072 (0.002) \\
& & HAC     & 0.287 (0.049) & 0.124 (0.004) & 0.939 (0.004) & 0.068 (0.002) \\
\cline{1-3}
\multirow{12}{*}{150} 
& \multirow{3}{*}{400} 
& Pop     & 0.902 (0.018) & 1.398 (0.000) & 0.999 (0.001) & 1.056 (0.000) \\
& & Plug-in & 0.381 (0.049) & 0.439 (0.041) & 0.951 (0.004) & 0.141 (0.013) \\
& & HAC     & 0.315 (0.044) & 0.414 (0.042) & 0.934 (0.003) & 0.129 (0.013) \\
\cline{2-3}
& \multirow{3}{*}{900}
& Pop     & 0.870 (0.022) & 0.939 (0.000) & 1.000 (0.000) & 0.710 (0.000) \\
& & Plug-in & 0.286 (0.034) & 0.235 (0.015) & 0.954 (0.003) & 0.097 (0.006) \\
& & HAC     & 0.246 (0.033) & 0.216 (0.016) & 0.936 (0.003) & 0.090 (0.006) \\
\cline{2-3}
& \multirow{3}{*}{1600}
& Pop     & 0.865 (0.025) & 0.708 (0.000) & 1.000 (0.000) & 0.535 (0.000) \\
& & Plug-in & 0.259 (0.038) & 0.161 (0.007) & 0.954 (0.003) & 0.079 (0.003) \\
& & HAC     & 0.232 (0.035) & 0.149 (0.007) & 0.938 (0.003) & 0.073 (0.003) \\
\cline{2-3}
& \multirow{3}{*}{2500}
& Pop     & 0.872 (0.027) & 0.567 (0.000) & 1.000 (0.000) & 0.429 (0.000) \\
& & Plug-in & 0.263 (0.039) & 0.127 (0.003) & 0.954 (0.002) & 0.069 (0.002) \\
& & HAC     & 0.242 (0.038) & 0.119 (0.003) & 0.940 (0.002) & 0.065 (0.001) \\
\bottomrule
\end{tabular}%
}
\caption{\textbf{Average coverage probabilities and coverage areas over the support sets $E$ and $E^c$ for the VAR(1) model at the almost-zero frequency $\omega=\pi/100$.} The reported values are averages over 200 repetitions (standard deviations in parentheses).}
\label{tab:sim_var1_almostzero_appendix}
\end{table}

\begin{table}[t]
\centering
\small
\renewcommand{\arraystretch}{1.05}
\resizebox{0.85\columnwidth}{!}{%
\begin{tabular}{ccccccc}
\toprule
$p$ & $n$ & Method 
& $\textrm{Avgcov}_{j}(E)$ 
& $\textrm{Area}_{j}(1,2)$ 
& $\textrm{Avgcov}_{j}(E^c)$ 
& $\textrm{Area}_{j}(1,p)$ \\
\midrule
\multirow{12}{*}{20} & \multirow{3}{*}{200} & Pop     & 0.924 (0.055) & 1.2369 & 1.000 (0.002) & 1.0826 \\
& & Plug-in & 0.537 (0.084) & 0.2512 & 0.929 (0.035) & 0.2777 \\
& & HAC     & 0.567 (0.064) & 0.2526 & 0.940 (0.025) & 0.2877 \\
\cline{2-3}
& \multirow{3}{*}{500} & Pop     & 0.922 (0.042) & 0.6548 & 0.998 (0.005) & 0.5731 \\
& & Plug-in & 0.615 (0.088) & 0.1666 & 0.926 (0.037) & 0.1769 \\
& & HAC     & 0.656 (0.066) & 0.1837 & 0.944 (0.020) & 0.1926 \\
\cline{2-3}
& \multirow{3}{*}{1000} & Pop     & 0.947 (0.032) & 0.4526 & 0.998 (0.006) & 0.3962 \\
& & Plug-in & 0.653 (0.086) & 0.1394 & 0.929 (0.034) & 0.1378 \\
& & HAC     & 0.700 (0.064) & 0.1490 & 0.944 (0.022) & 0.1476 \\
\cline{2-3}
& \multirow{3}{*}{2000} & Pop     & 0.965 (0.029) & 0.2757 & 0.998 (0.006) & 0.2413 \\
& & Plug-in & 0.755 (0.076) & 0.1016 & 0.936 (0.035) & 0.0987 \\
& & HAC     & 0.770 (0.060) & 0.1082 & 0.946 (0.025) & 0.1048 \\
\cline{1-3}
\multirow{12}{*}{50} & \multirow{3}{*}{200} & Pop     & 0.970 (0.028) & 1.2280 & 1.000 (0.001) & 1.0748 \\
& & Plug-in & 0.535 (0.095) & 0.1928 & 0.930 (0.035) & 0.2102 \\
& & HAC     & 0.554 (0.068) & 0.2051 & 0.939 (0.023) & 0.2296 \\
\cline{2-3}
& \multirow{3}{*}{500} & Pop     & 0.978 (0.023) & 0.6760 & 1.000 (0.001) & 0.5916 \\
& & Plug-in & 0.610 (0.098) & 0.1363 & 0.924 (0.037) & 0.1458 \\
& & HAC     & 0.644 (0.091) & 0.1554 & 0.943 (0.023) & 0.1655 \\
\cline{2-3}
& \multirow{3}{*}{1000} & Pop     & 0.983 (0.017) & 0.4758 & 1.000 (0.002) & 0.4165 \\
& & Plug-in & 0.651 (0.102) & 0.1236 & 0.932 (0.036) & 0.1192 \\
& & HAC     & 0.694 (0.084) & 0.1275 & 0.945 (0.024) & 0.1245 \\
\cline{2-3}
& \multirow{3}{*}{2000} & Pop     & 0.977 (0.021) & 0.2662 & 1.000 (0.003) & 0.2330 \\
& & Plug-in & 0.734 (0.084) & 0.0889 & 0.938 (0.034) & 0.0884 \\
& & HAC     & 0.745 (0.067) & 0.0949 & 0.948 (0.022) & 0.0936 \\
\cline{1-3}
\multirow{12}{*}{100} & \multirow{3}{*}{200} & Pop     & 0.975 (0.026) & 1.2160 & 1.000 (0.001) & 1.0643 \\
& & Plug-in & 0.509 (0.113) & 0.1770 & 0.930 (0.036) & 0.1872 \\
& & HAC     & 0.518 (0.082) & 0.1891 & 0.940 (0.024) & 0.2037 \\
\cline{2-3}
& \multirow{3}{*}{500} & Pop     & 0.979 (0.021) & 0.6508 & 1.000 (0.001) & 0.5696 \\
& & Plug-in & 0.572 (0.121) & 0.1200 & 0.925 (0.037) & 0.1304 \\
& & HAC     & 0.618 (0.110) & 0.1384 & 0.944 (0.023) & 0.1491 \\
\cline{2-3}
& \multirow{3}{*}{1000} & Pop     & 0.985 (0.018) & 0.4245 & 1.000 (0.001) & 0.3716 \\
& & Plug-in & 0.640 (0.119) & 0.1057 & 0.931 (0.035) & 0.1049 \\
& & HAC     & 0.675 (0.105) & 0.1145 & 0.946 (0.023) & 0.1141 \\
\cline{2-3}
& \multirow{3}{*}{2000} & Pop     & 0.978 (0.022) & 0.2471 & 1.000 (0.002) & 0.2163 \\
& & Plug-in & 0.715 (0.097) & 0.0794 & 0.936 (0.035) & 0.0773 \\
& & HAC     & 0.732 (0.082) & 0.0862 & 0.948 (0.022) & 0.0849 \\
\bottomrule
\end{tabular}%
}
\caption{\textbf{Average coverage probabilities over the support sets $E$ and $E^c$ for the VMA(3) model at the non-zero frequency $\omega=\pi/2$.} The reported coverage values are averages over 200 repetitions (standard deviations in parentheses). The area columns report the corresponding average confidence-region areas for the representative entries $(1,2)\in E$ and $(1,p)\in E^c$.}
\label{tab:sim_vma3_coverage_area}

\end{table}

\end{document}